\documentclass[aps, a4paper, reprint, preprintnumbers]{revtex4-2}

\usepackage{amsmath, amssymb, amsfonts, physics}
\usepackage{hyperref}
\hypersetup{
	colorlinks=true,
	linkcolor=blue
}
\usepackage{graphicx}
\newtheorem{theorem}{Theorem}
\usepackage{blindtext}
\usepackage{xcolor}

\begin{document}
\title{Multi-invariants in stabilizer states}
\date{\today}
\author{Sriram Akella}
\email{sriram.akella@tifr.res.in}
\author{Abhijit Gadde}
\email{abhijit.gadde@tifr.res.in}
\author{Jay Pandey}
\email{jay.pandey@tifr.res.in}
\affiliation{Department of Theoretical Physics, Tata Institute of Fundamental Research, Homi Bhabha Road, Mumbai  400005, India.}
\begin{abstract}
Multipartite entanglement is a natural generalization of bipartite entanglement, but is relatively poorly understood. In this paper, we develop tools to calculate a class of multipartite entanglement measures - known as multi-invariants - for stabilizer states. We give an efficient numerical algorithm that computes multi-invariants for stabilizer states. For tripartite stabilizer states, we also obtain an explicit formula for any multi-invariant using the GHZ-extraction theorem. We then present a counting argument that calculates any Coxeter multi-invariant of a $q$-partite stabilizer state. We conjecture a closed form expression for the same. We uncover hints of an interesting connection between multi-invariants, stabilizer states and topology. We show how our formulas are further simplified for a restricted class of stabilizer states that appear as ground states of interesting models like the toric code and the X-cube model.    
\end{abstract}
\preprint{TIFR/TH/26-4}
 \maketitle

\section{Introduction}
Entanglement is a phenomenon with no counterpart in classical systems \cite{Horodecki:2009zz}. It is a valuable resource \cite{Chitambar:2018rnj} in various quantum tasks including quantum teleportation \cite{Bennett:1992tv}, superdense coding \cite{Bennett:1992zzb}, quantum key distribution \cite{BENNETT20147, Ekert:1991zz}, and quantum algorithms \cite{Kendon:2006wyl}. Beyond these information-theoretic and computational applications, studying quantum entanglement has also enhanced our understanding of the long-range correlations in condensed matter systems \cite{Laflorencie:2015eck}. In the context of the AdS/CFT correspondence \cite{Hubeny:2014bla}, entanglement has helped us understand the local aspects of the encoding of the gravitational degrees of freedom into conformal field theory. In particular, it has given insights into the so-called entanglement wedge reconstruction \cite{Cotler:2017erl}, the black hole information paradox \cite{Almheiri:2020cfm}, subregion-subregion duality \cite{Leutheusser:2022bgi}, and various other features of holography \cite{Nishioka:2009un}.

Discussion of entanglement in the literature is mostly centered around bipartite systems. In such systems, the Hilbert space is factorized into two as $\mathcal{H} = \mathcal{H}_A \otimes \mathcal{H}_B$. A state $\ket{\psi} \in \mathcal{H}$ is entangled if it cannot be written as a simple product state of the form $\ket{\psi_A} \otimes \ket{\psi_B}$, where $\ket{\psi_A} \in \mathcal{H}_A$ and $\ket{\psi_B} \in \mathcal{H}_B$. For pure bipartite states, the von Neumann entropy (or entanglement entropy), defined as $\mathcal{S}(\ket{\psi}) = -\Tr_A \rho_A \log \rho_A = - \Tr_B \rho_B \log \rho_B$, is a quantitative measure of entanglement between subsystems $A$ and $B$. Here, $\rho_A$ and $\rho_B$ are the reduced density matrices obtained by tracing out the subsystem $B$ and $A$, respectively.

In this paper, we are interested in studying multipartite entanglement \cite{bengtsson2016briefintroductionmultipartiteentanglement}. This is the entanglement within the state living in ${\cal H}$ that is factorized into more than two factors. 
Quantifying multipartite entanglement is a difficult task due to the existence of qualitatively different types of such entanglement.
In the simplest multipartite scenario involving three qubits, there are two distinct entanglement classes \cite{Dur:2000zz}: the GHZ and the W class. For four qubits, the situation is more complex with nine entanglement classes, some of which form continuous families \cite{Verstraete_2002}. As the number of parties or the dimensionality of Hilbert space increases, the structure of multipartite entanglement becomes increasingly intricate \cite{Sauerwein_2018}. To understand this complexity, we require robust quantitative measures of multipartite entanglement \cite{Horodecki:2024bgc, Ma:2023ecg, Szalay:2015vrx}. We use a class of local unitary invariants, called multi-invariants \cite{Gadde:2024taa}, for this purpose. Multi-invariants have been useful in construction of entanglement monotones \cite{Gadde:2025csh} as well as in providing a geometric characterization of multipartite entanglement in holographic systems \cite{Gadde:2024taa}.

As we will explain below, each $q$-partite multi-invariant is labeled by a set of $q$ permutations $(\sigma_1, \dots, \sigma_q)$, up to the equivalence $(\sigma_1, \dots, \sigma_q)\sim \tau \cdot (\sigma_1, \dots, \sigma_q)$, that belong to the  symmetric group ${S}_n$ for some $n$. Here $n$ is called the \emph{replica number}. An immediate problem that confronts us is a lack of technology to compute multi-invariants for arbitrary states. This is because we need to sum over $nq$ tensor legs where each leg carries a Hilbert space dimension worth of sums \cite{hillar2013most}. Even working with qubits with modest values of $n = 9$ and $q = 3$, there are $2^{27} = 134217728$ sums to be performed.

In this paper, we focus on a class of quantum states known as stabilizer states \cite{Gottesman:1997zz} and develop specialized tools to compute their multi-invariants. Stabilizer states are well-suited for such studies due to the Gottesman-Knill theorem \cite{Gottesman:1998hu}  which allows them to be efficiently stored and manipulated on classical computers \cite{Aaronson:2004xuh}. Despite their simplicity, stabilizer states exhibit a rich structure of multipartite entanglement \cite{Hein:2004zjp, Fattal:2004frh}, making them ideal candidates for the study of multi-invariants.
Moreover, stabilizer states play a foundational role in various quantum technologies, including CSS quantum error-correcting codes \cite{Calderbank:1995dw, Steane:1995vv}, surface codes \cite{Fowler_2012}, color codes \cite{Landahl:2011den, Fowler_2011}, one-way quantum computation \cite{raussendorf2003measurement, raussendorf2001one}, and numerous condensed matter systems based on stabilizer Hamiltonians \cite{Terhal:2013vbm}. Stabilizer states also appear in abelian Chern-Simons theory \cite{Salton:2016qpp, Balasubramanian:2016sro, Balasubramanian:2018por, Yuan:2025dgx}, further emphasizing their broad relevance. Given their importance, understanding multipartite entanglement of stabilizer states is crucial, and we make progress in this direction in the present work.

The remainder of this paper is organized as follows. In Section \ref{sec:multi-invariants}, we review the construction of multi-invariants and define the R\'enyi multi-entropy, introduced in \cite{Gadde:2022cqi}, as a prototypical example. In Section \ref{sec:stabilizer}, we provide an overview of stabilizer states, their connection to graph states, and review their bipartite entanglement. In Section \ref{sec:graph-states}, we demonstrate how the computation of multi-invariants for stabilizer states reduces to an inner product problem, using the example of R\'enyi multi-entropy. We also introduce a polynomial-time numerical algorithm for computing inner products of stabilizer states. Section \ref{sec:tripartite-stabilizer-states} focuses on tripartite stabilizer states, where we show that their multi-invariants take a simplified form, leveraging a theorem proven in \cite{Bravyi_2006}. This discussion naturally leads to an exploration of multi-invariants derived from Coxeter groups, which also appear in recent works \cite{Gadde:2024taa, Gadde:2025csh, Gadde:2025ybn}. We make a powerful conjecture about these Coxeter multi-invariants for stabilizer states. We offer a counting argument  for this conjecture and check it in the case of tripartite stabilizer states. This is followed by Section \ref{sec:restricted} where we present a simplification of the Coxeter multi-invariants for a restricted class of stabilizer states that appear as the ground states of spin systems like the toric code and the X-cube model. 
We conclude in Section \ref{sec:discussion} with a summary of our findings and suggestions for future research.

\section{Multi-invariants}\label{sec:multi-invariants}

Suppose $\ket{\psi}$ is a multipartite pure state shared between $q$ parties: $\ket{\psi} \in \mathcal{H}_1 \otimes \dots \otimes \mathcal{H}_q$, where each Hilbert space $\mathcal{H}_i$ is treated as a distinct party. A multi-invariant ${\cal Z}$ is a local unitary invariant homogeneous polynomial of components of $|\psi\rangle$ and their complex conjugates that is multiplicative. Let us elaborate on these properties:
\begin{enumerate}
    \item \textbf{LU invariance.} \\
    $\mathcal{Z}$ is invariant under a local unitary transformation, i.e., $\mathcal{Z}\left(\left(U_1 \otimes \dots \otimes U_q\right) \ket{\psi} \right) = \mathcal{Z} \left(\ket{\psi} \right)$, where each $U_a: \mathcal{H}_a \to \mathcal{H}_a$ is unitary.
    \item \textbf{Homogeneity.} \\
    ${\mathcal Z}(c|\psi\rangle)=|c|^{2n}{\mathcal Z}(|\psi\rangle)$. Here $n$ is known as the replica number. 
    \item \textbf{Multiplicativity.} \\
    ${\mathcal Z}(|\psi\rangle_1\otimes |\psi\rangle_2)={\mathcal Z}(|\psi\rangle_1){\mathcal Z}(|\psi\rangle_2)$. Here both $|\psi_1\rangle$ and $|\psi_2\rangle$ are $q$-partite states. 
\end{enumerate}
In \cite{Gadde:2025csh}, a large class of multipartite entanglement monotones was constructed using multi-invariants. It was conjectured in \cite{Gadde:2025csh,Gadde:2025ybn} that $\hat \nu := 1-{\cal Z}^{1/n}$ is an entanglement monotone where ${\cal Z}$ is a ``symmetric'' multi-invariant based on a finite Coxeter group. The conjecture was proved for all but six exceptional finite Coxeter groups \cite{Gadde:2025ybn}. In this paper, we will work with general multi-invariants.

\begin{figure}
\centering
	\includegraphics[width = 0.75\linewidth]{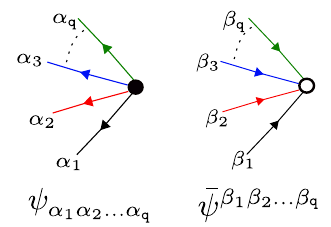}
	\caption{Graphical notation for the coefficients of $\psi$ and the conjugate $\bar{\psi}$.}
	\label{fig:psi}
\end{figure}

Multi-invariants of a $q$-party state $\ket{\psi}$ are constructed as follows. Let $d_a = \dim(\mathcal{H}_a)$ be the dimensions of the various Hilbert spaces, where $a = 1,\ldots,q$, and let $\{\ket{\alpha_a}\}$ be a basis for each $\mathcal{H}_a$, where $\alpha_a = 1, \dots, d_a$. Writing the state $\ket{\psi}$ in the factorized basis as
\begin{equation}\label{eq:1.1}
 \ket{\psi} = \sum_{\alpha_1 = 1}^{d_1} \dots \sum_{\alpha_q = 1}^{d_q} \psi_{\alpha_1 \dots \alpha_q} \ket{\alpha_1} \otimes \dots \otimes \ket{\alpha_q},
\end{equation}
we interpret the coefficients $\psi_{\alpha_1 \dots \alpha_q}$ and its complex conjugate graphically as shown in figure \ref{fig:psi}. The coefficient $\psi_{\alpha_1 \dots \alpha_q}$ is represented by a dark vertex with $q$ outgoing colored edges carrying labels $\alpha_1, \dots, \alpha_q$. The coefficient $\bar{\psi}^{\beta_1 \dots \beta_q}$ of $\bra{\psi}$ is represented by a light vertex with $q$ incoming colored edges carrying labels $\beta_1, \dots, \beta_q$.

To construct local unitary invariants, we take $n$ copies of $\psi$ and contract the outgoing legs with the incoming legs of $n$ copies of $\bar{\psi}$. As an example, consider the second R\'enyi entropy of a bipartite state $\psi_{\alpha \beta}$:
\begin{equation}
    \includegraphics[width = \linewidth]{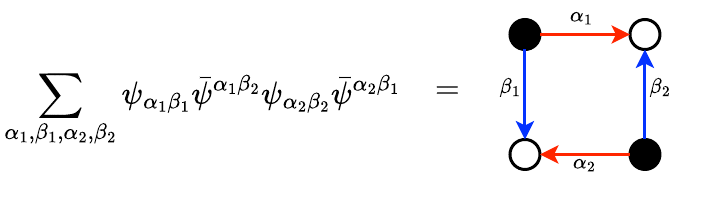}
\end{equation}
In this example, the replica number is $n = 2$. More generally, when we have $n$ replicas of a $q$-partite state $\psi$, there are $nq$ outgoing legs which need to be matched with $nq$ incoming legs of $n$ replicas of $\bar{\psi}$ to produce a local unitary invariant. If the state $\ket{\psi}$ is fully factorized, then any multi-invariant evaluates to $1$ (assuming $\ket{\psi}$ is normalized).

\begin{figure}
	\centering
	\includegraphics[width = 0.15\textwidth]{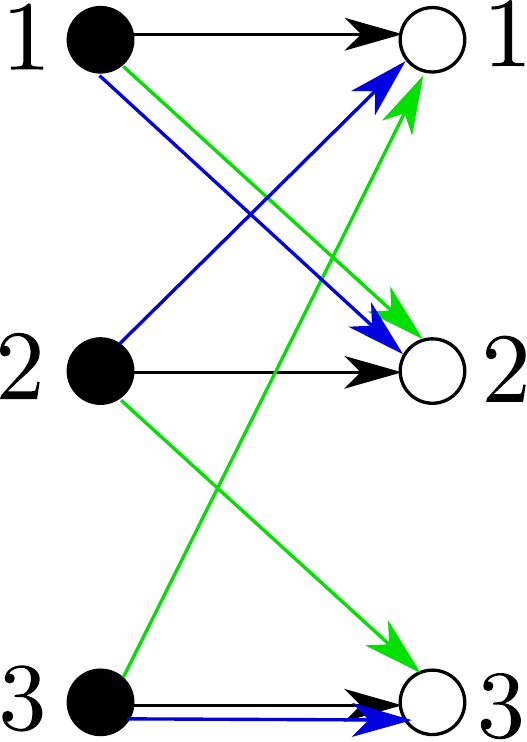}
	\caption{The multi-invariant corresponding to $\sigma_1 = (1)(2)(3), \sigma_2 = (123), \sigma_3 = (12)(3)$.}
	\label{fig:perms}
\end{figure}

Alternatively, we can also define a multi-invariant by associating a permutation $\sigma_a \in {S}_n$ to each party $a = 1, \dots, q$, where ${S}_n$ is the symmetric group on $n$ letters. The permutation $\sigma_a$ tells us how to match the legs of party $a$.

As an example, suppose we have a tripartite state with $n = 3$ replicas and the three permutations $\sigma_1 = (1)(2)(3)$, $\sigma_2 = (123)$, and  $\sigma_3 = (12)(3)$.
The corresponding multi-invariant is shown in figure \ref{fig:perms}. The black legs, corresponding to party $1$, follow the permutation $\sigma_1 = (1)(2)(3)$. Therefore we connect the first copy of $\psi$ with the first copy of $\bar{\psi}$, etc. The green legs, corresponding to party $2$, follow the permutation $\sigma_2 = (123)$. Therefore we connect the first copy of $\psi$ to the second copy of $\bar{\psi}$, second copy of $\psi$ to the third copy of $\bar{\psi}$, and the third copy of $\psi$ to the first copy of $\bar{\psi}$. Similarly for the blue legs that follow the permutation $\sigma_3 = (12)(3)$. 

The multi-invariant can then be defined as a function of the permutation tuple $(\sigma_1,\ldots, \sigma_q)$
\begin{align}
	{\cal Z}(\sigma_1,\ldots, \sigma_q)=\langle \psi|^{\otimes n} \sigma_1\otimes \ldots \otimes \sigma_q|\psi\rangle^{\otimes n}.
\end{align}
This labeling is redundant because, the bras and the kets can be independently relabeled. This gives the relation,
\begin{equation}\label{eq:relabel}
\begin{split}
 	\mathcal{Z}(\sigma_1, \dots, \sigma_q) & = \mathcal{Z}(g\sigma_1 , \dots, g\sigma_q) \\
      & = \mathcal{Z}(\sigma_1 h, \dots, \sigma_q h),
 \end{split}
 \end{equation}
 where $g$ and $h$ are elements of the permutation group ${S}_n$. We appeal to figure \ref{fig:perms} for a quick explanation. Left(right)-multiplication by $g$ is equivalent to relabeling the dark (light) vertices by $g$. This relabeling does not change the value of the multi-invariant.

\begin{figure}
    \centering
    \includegraphics[width=0.75\linewidth]{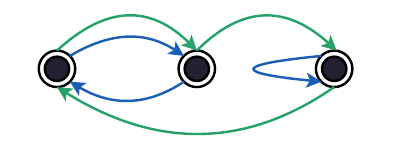}
    \caption{Multi-invariant from figure \ref{fig:perms} after tracing out party $1$. Each vertex stands for a reduced density matrix on the remaining $(q-1)$ parties.}
    \label{fig:reduced-face}
\end{figure}

We may use this redundancy to set $\sigma_q = (1)(2)\dots(n)$; the identity permutation.  In our graphical notation, the edges corresponding to party $q$ go straight from the ket to the bra of the same replica label (see the black edge in figure \ref{fig:perms}, for example.)
We simplify our graphical notation by removing this edge of party $q$ and putting the bra and ket on top of each other.
Our graph now has $n$ vertices, and each vertex has $(q-1)$ incoming and $(q-1)$ outgoing edges, see figure \ref{fig:reduced-face} for an example. Therefore, the interpretation of a vertex is as the $(q-1)$-partite reduced density matrix obtained after tracing out party $q$ from the $q$-partite state $\ket{\psi}$. In terms of an equation,
\begin{equation}
    \rho_{\alpha_1 \dots \alpha_{q-1}}^{\beta_1 \dots \beta_{q-1}} = \sum_{\alpha_q} \psi_{\alpha_1 \dots \alpha_{q-1} \alpha_q} \bar{\psi}^{\beta_1 \dots \beta_{q-1} \alpha_q}.
\end{equation}
The $(q-1)$ outgoing legs are interpreted as the ket indices $\alpha_1, \dots, \alpha_{q-1}$ and the incoming legs as bra indices $\beta_1, \dots, \beta_{q-1}$.

\subsection{Multi-entropy} 
\begin{figure}
	\includegraphics[width = \linewidth]{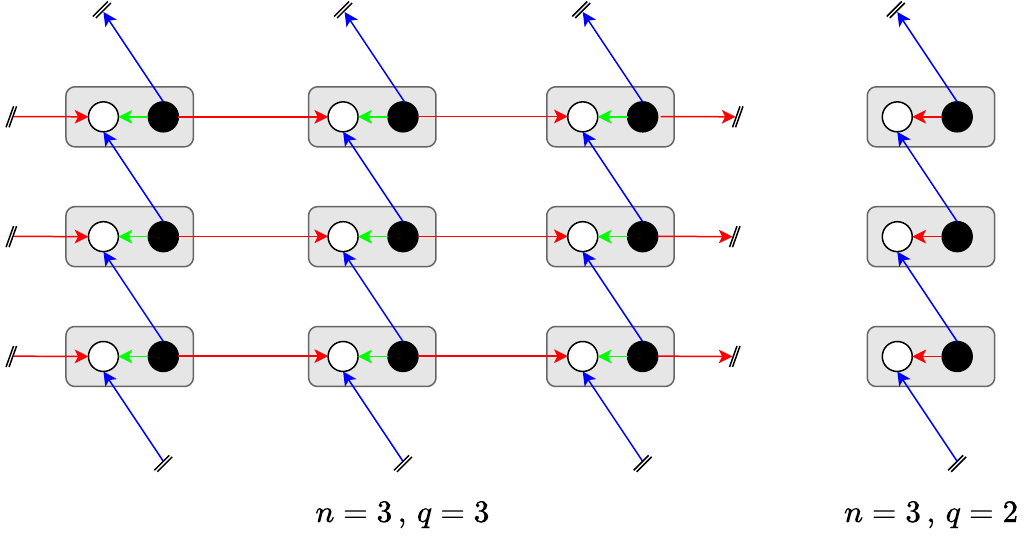}
	\caption{The multi-invariant graph that computes the multi-entropy for $n = 3, q = 3$, and $n = 3, q = 2$ respectively. }
	\label{fig:multi-entropy-examples}
\end{figure}

In this subsection, we introduce an example of a multi-invariant called the multi-entropy \cite{Gadde:2022cqi}.
Given a $q$-party state $\ket{\psi} \in \mathcal{H}_1 \otimes \dots \otimes \mathcal{H}_q$, we define the multi-invariant ${\cal Z}_n^{(q)}$  as follows. Start with a $q-1$-dimensional cubic lattice with $n$ sites in each of the $q-1$ directions and periodic boundary conditions. The sites of this lattice can be labeled by vectors in $\mathbb{Z}_n^{q-1}$. A copy of $\psi$ with $q$ outgoing legs and a copy of $\bar{\psi}$ with $q$ incoming legs live at each of the $n^{q-1}$ sites of the cubic lattice. For $a<q$, the outgoing leg of the $a^{\text{th}}$ party of the $\psi$ at the location $\vec{r} \in \mathbb{Z}_n^{q-1}$ is connected to the incoming $a^{\text{th}}$ leg of the $\bar{\psi}$ at the site
\begin{equation}\label{eq:multientropy-perms}
	\sigma_a(\vec{r}) = (r_1, \dots, r_a + 1, \dots, r_{q-1}),
\end{equation}
where the addition $r_a + 1$ is modulo $n$. As for the $q$-th party, the $q$-th leg of $\psi$ at location $\vec r$ is connected to the $q$-th leg of $\bar \psi$ at the same location i.e. the permutation element associated to the $q$-th party is identity. In defining  ${\cal Z}_n^{(q)}$, we have treated $q$-th party differently from all the other parties, however, thanks to the redundancy \eqref{eq:relabel} in labeling with permutation tuple, this multi-invariant is in fact symmetric in all parties. This is made clear in the two examples  shown in figure \ref{fig:multi-entropy-examples}.

The R\'enyi multi-entropies are defined as 
\begin{equation}\label{eq:Renyi-multis}
	\mathcal{E}_n^{(q)\text{ME}} = \frac{1}{(1-n)} \log \mathcal{Z}_n^{(q)},
\end{equation}
where $\mathcal{Z}_n^{(q)}$ is the multi-invariant defined above. The multi-entropy is defined as the $n \to 1$ limit of the R\'enyi multi-entropy. The R\'enyi multi-entropy reduces to the usual R\'enyi entropies when the number of parties, $q = 2$.

We now turn our attention to stabilizer states: the arena where we will evaluate these multi-invariants.

\section{Stabilizer states}\label{sec:stabilizer}
Representing a generic quantum state of $n$ qubits on a classical computer requires $O(2^n)$ complex numbers. However, a class of states, called stabilizer states, are much cheaper to manipulate on a classical computer \cite{Gottesman:1997zz, Gottesman:1998hu}. They are defined as follows.

Recall that the Pauli group is the group generated by the Pauli matrices. The Pauli group on $n$ qubits, denoted by $\mathcal{P}_n$, contains all the $n$-fold tensor products of the elements of the Pauli group $\mathcal{P}_1$. An $n$-qubit quantum state $\ket{\psi}$ is a \emph{stabilizer state} if there is a set of $n$ mutually commuting elements $\{g_1, \dots, g_n\}$ in the Pauli group $\mathcal{P}_n$, such that
\begin{equation}
 g_i \ket{\psi} = \ket{\psi}.
\end{equation}
The subgroup $G = \langle g_1, \dots, g_n \rangle \subset \mathcal{P}_n$ generated by these generators is called the \emph{stabilizer group} of the state $\ket{\psi}$. Given that $g_i^2=1$, the stabilizer group consists of elements of the form $g_1^{\alpha_1}\ldots g_n^{\alpha_n}$ where $\alpha_i \in \{0,1\}$.
The information about a stabilizer state $\ket{\psi}$ is equivalently found in its stabilizer group $G$.

As an example, consider the Bell pair: $\ket{\psi} = 1/\sqrt{2}(\ket{00} + \ket{11})$. It is stabilized by the mutually commuting elements $g_1 = Z\otimes Z$ and $g_2 = X \otimes X$, where $X$ and $Z$ are the usual Pauli operators defined by $X\ket{0} = \ket{1}$, $X \ket{1} = \ket{0}$,  $Z\ket{0} = \ket{0}$, and $Z\ket{1} = - \ket{1}$. The corresponding stabilizer group is $G =$ $\{$$g_1^0g_2^0=I \otimes I$$,\, g_1^1 g_2^0=Z \otimes Z$$,\, g_1^0 g_2^1= X\otimes X$$,\, g_1^1g_2^1=- Y \otimes Y$$ \}$.

As an example of a state that is \emph{not} a stabilizer state, consider the W-state defined on three qubits,
\begin{equation}
	\ket{\text{W}} = \frac{1}{\sqrt{3}} \left(\ket{001} + \ket{010} + \ket{100} \right).
\end{equation}
No element of the Pauli group on three qubits, except the identity, stabilizes the above state. Hence the W state is not a stabilizer state.

\subsection{Entanglement Entropy}
Having defined stabilizer states, let's now discuss how to compute their entanglement entropies. Owing to the stabilizer structure, there is a simple formula we can write down for the entanglement entropy.

Suppose $\ket{\psi}$ is a stabilizer state on $n$ qubits with a stabilizer group $G$. It can be written in terms of the stabilizer group elements as,
\begin{equation}\label{eq:density}
\ket{\psi}\bra{\psi} = \frac{1}{2^n} \sum_{g \in G} g.
\end{equation}
If we partition the $n$ qubits into two parties $A$, and $B$, then each element $g \in G$ can be written as $g_A \otimes g_B$. Here $g_A$ acts only on qubits in party $A$, and $g_B$ acts only on qubits in $B$. Tracing out party $A$, for instance, gives
\begin{equation}\label{eq:1.12}
 \rho_{B} =  \frac{1}{2^{|B|}} \sum_{g \in G_{B}} g,
\end{equation}
where $|B|$ is the number of qubits in party $B$, and $G_{B}$ is the subgroup of all elements of $G$ which act as identity on $A$. This is because the only element of the Pauli group that has non-vanishing trace is identity.

From this expression for the reduced density matrix, we can compute the R\'enyi entropies of $\rho_B$:
\begin{equation}
    \mathcal{S}_m^{(2)} = \frac{1}{1-m} \log_2 \Tr \rho_B^m = |B| - \log_2 |G_B|,
\end{equation}
where $|G_B|$ is the cardinality of the subgroup $G_B$. Repeating the same calculation with $\rho_A$, we get $\mathcal{S}_n^{(2)} = |A| - \log_2 |G_A|$. Although not obvious in this form, these two quantities are equal \cite{Fattal:2004frh}, and we can represent the entanglement entropy by the symmetrized formula:
\begin{equation}\label{eq:ent-ent}
    \mathcal{S}_m^{(2)} = \frac{1}{2} \log_2 \left(\frac{|G|}{|G_A| |G_B|}\right).
\end{equation}
Here, $|G| = 2^n$ is the cardinality of the full stabilizer group. The entanglement entropy is the same as any other R\'enyi entropy as $\mathcal{S}_m^{(2)}$ does not depend on $m$. 

\subsection{Graph States}
Within stabilizer states, there is a special class of states called \emph{graph states} \cite{Hein:2006uvf}. A graph $G = (V, E)$ comprises a set of vertices, $V$, and a set of edges $E$ which are unordered pairs of vertices: $E \subseteq \{ (x, y): x, y \in V, x\neq y\}$. Given a graph $G = (V, E)$, define the corresponding graph state $\ket{G}$ as the state stabilized by the following elements of the Pauli group:
\begin{equation}
g_a = X_a \prod_{b: (a, b) \in E} Z_b.\label{eq:GraphStateDef}
\end{equation}
It's easy to check that this forms a commuting set, and therefore defines a stabilizer state.

\begin{figure}
    \centering
    \includegraphics[width=0.5\linewidth]{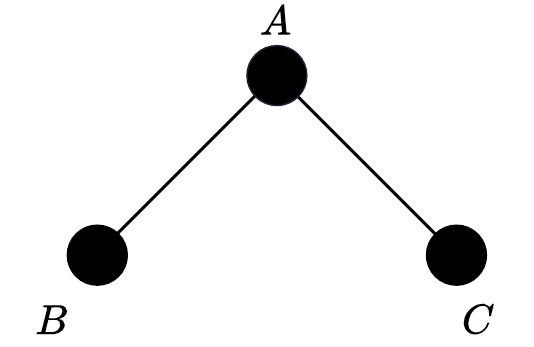}
    \caption{A simple graph $G$ with three vertices $V = \{A, B, C\}$ and two edges $ E = \{(A,B), (B,C)\}$.}
    \label{fig:ghz-graph}
\end{figure}

As an example, consider the graph shown in figure \ref{fig:ghz-graph}. The corresponding graph state, by definition, is the state stabilized by $g_A = X_A Z_B Z_C$, $g_B = Z_A X_B I_C$, $g_C = Z_A I_B X_C$, where $I$ is the $2 \times 2$ identity operator and the subscript stands for the site where the operator acts.

Although graph states are a special class of stabilizer states, every stabilizer state is \emph{local Clifford-equivalent} to a graph state \cite{Nest:2004khg}. To understand what local Clifford-equivalence means, we will have to understand how stabilizer states transform under a unitary transformation.

Under a unitary transformation $U$, a stabilizer state $\ket{\psi}$ with stabilizer group $G$ transforms as $\ket{\psi'} = U \ket{\psi}$. The new stabilizer group, therefore, is
\begin{equation}\label{eq:2.7}
  G' = \{ U g U^{\dagger}: g \in G\}.
\end{equation}
Although $g$ is an element of the Pauli group $\mathcal{P}_n$, the new element $U g U^{\dagger}$ is, in general, not in $\mathcal{P}_n$. However, the set of unitaries $\mathcal{C}_n =  \{U: U g U^{\dagger} \in \mathcal{P}_n\ \text{for all } g \in \mathcal{P}_n\}$, form a group called the \emph{Clifford group}. The stabilizer states $\ket{\psi}$ and $\ket{\psi'}$ are said to be \emph{Clifford-equivalent} if an element of the Clifford group takes $\ket{\psi}$ to $\ket{\psi'}$. A local Clifford unitary is a local unitary
that also happens to be an element of the Clifford group $\mathcal{C}_n$.

Since multi-invariants are local unitary invariant, we can restrict our attention to graph states. We develop an efficient algorithm to compute multi-invariants for graph states which we'll describe in detail in the following sections.

To this end, it is convenient to use an alternative but equivalent definition of graph states:  Given a graph $G = (V, E)$ with $n$ vertices, imagine a qubit living on each vertex. We begin with the state
\begin{equation}
 \ket{\text{start}} = \ket{+}^{\otimes n},
\end{equation}
where $\ket{+}$ is the eigenstate of the Pauli $X$ operator with eigenvalue $+1$, i.e., $ X\ket{+} = \ket{+}$.
For every edge of the graph $(a, b) \in E$, define $U_{(a, b)}$ as the controlled-$Z$ operation on the two qubits located at vertices $a$ and $b$:
\begin{equation}
\begin{split}
 U_{(a, b)} & = \ket{0}\bra{0}_a \otimes I_b + \ket{1}\bra{1}_a \otimes Z_b \\
 & = \left(\frac{1+Z_a}{2}\right)\otimes I_b + \left(\frac{1 - Z_a}{2}\right)\otimes Z_b.
 \end{split}
\end{equation}
Here $\ket{0}$ and $\ket{1}$ are eigenvectors of $Z$ with eigenvalues $+1$ and $-1$ respectively. This operator is equal to $\mathrm{diag}(1,1,1,-1)$ in the computational basis and is symmetric under $a \leftrightarrow b$.
The graph state $\ket{G}$ is then
\begin{equation}\label{eq:2.6}
 \ket{G} = \prod_{(a, b) \in E} U_{(a, b)}\,\ket{\text{start}}.
\end{equation}
The  two $U$'s commute even when they share a vertex. As a result, the order of these unitaries does not matter, making the above expression unambiguous.

\section{Multi-invariants in graph states}\label{sec:graph-states}
In this section, we show that calculating a multi-invariant for a graph state can be mapped to the computation of an inner product. Then we present an efficient numerical algorithm to compute the inner product.

\subsection{Multi-invariants as inner products}
\begin{figure*}
\includegraphics[width = \linewidth]{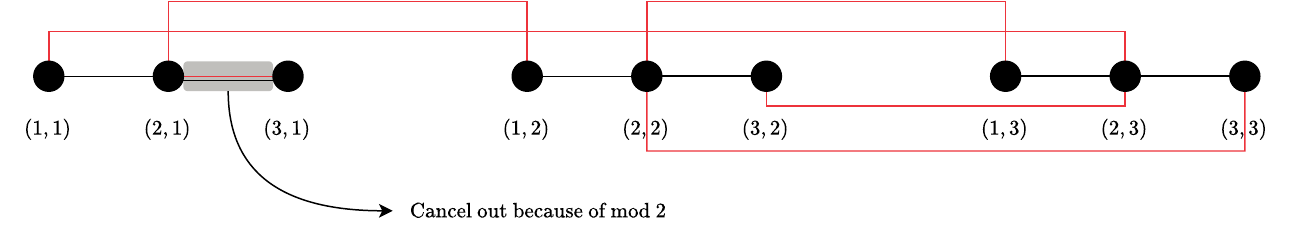}
\caption{The big graph $\mathcal{G}_n$ that computes the multi-invariant in figure \ref{fig:perms} for the GHZ graph state in figure \ref{fig:ghz-graph}.}
\label{fig:big-graph}
\end{figure*}

In this subsection, we will show that any multi-invariant $\mathcal{Z}(\sigma_1, \dots, \sigma_q)$ of replica number $n$ is given by an inner product of the form
\begin{equation}\label{eq:inner-product}
    \mathcal{Z}(\sigma_1, \dots, \sigma_q) = \langle + \dots +| \mathcal{G}_n \rangle
\end{equation}
for a graph state $\ket{G}$.
Here $\mathcal{G}_n$ is a \emph{big} graph state obtained from $G$ via the following procedure. It has $n|V|$ vertices. It is convenient to label its vertices by the tuple $(a, i)$ where $a$ is a vertex of the graph $G$ and $i$ ranges from $1$ to $n$. If $\gamma_{ab}$ is the \emph{adjacency matrix} of the graph $G$ then the adjacency matrix of the big graph $\mathcal{G}_n$ is
\begin{equation}\label{eq:big-adj}
    \Gamma_{(a, i), (b, j)} = \gamma_{ab} \left(\delta_{i, j} + \delta_{\sigma_a(i), \sigma_b(j)}\right) \mod 2
\end{equation}
where $\sigma_a$ is the permutation associated with the party to which vertex $a$ belongs. As an example, consider the multi-invariant from figure \ref{fig:perms} and the GHZ graph in figure \ref{fig:ghz-graph} where we treat each vertex as a distinct party. The resulting \emph{big graph} is shown in figure \ref{fig:big-graph}.

Multi-invariants simplify because of the operational definition of graph states in equation  (\ref{eq:2.6}). We can think of the multi-invariant $\mathcal{Z}$ of a graph state $\ket{G}$ as the expectation value of a bunch of permutation operators acting on a tensor product of $n$ copies of $\ket{G}$. We can schematically write this as
\begin{equation*}
\mathcal{Z}(\sigma_1, \dots, \sigma_q) = \langle G|^{\otimes n} (\sigma_1\otimes  \ldots\otimes  \sigma_q) | G \rangle^{\otimes n}.
\end{equation*}
Using equation  (\ref{eq:2.6}) we can replace $\bra{G}^{\otimes n}$ with $\bra{+ \dots +} U \dots U$, commute the $U$'s through the permutation operators and make them act on $\ket{G}^{\otimes n}$. This adds extra edges to $n$ copies of $G$. The permutation operators can then be made to act on $\bra{+ \dots +}$ which remains invariant. We are therefore left with equation  (\ref{eq:inner-product}). The extra edges added to $\ket{G}^{\otimes n}$ are given by the second term in equation  (\ref{eq:big-adj}) which are shown in red in figure \ref{fig:big-graph}.

\subsection{Inner product algorithm}
Having reduced the problem to an inner product, we now develop an efficient numerical algorithm. The way we compute the inner product in equation  \eqref{eq:inner-product} is by acting with the projector $P_{x, +}^{(a)} = \ket{+}\bra{+}^{(a)}$ iteratively for every vertex $a$ in $\mathcal{G}_n$. What's left in the end is the inner product we want. Graph states have nice properties under projective measurements \cite{Hein:2004zjp} which we now review.

\subsubsection{Local Pauli Measurements}
Consider the following projection operators that act at a vertex $a \in V$ of a given graph $G = (V, E)$:
\begin{equation}
 P_{i, \pm}^{(a)} = \frac{1}{2}\left(1 + \sigma_i^{(a)}\right) = \ket{i, \pm}\bra{i, \pm}^{(a)},
\end{equation}
where $i = x, y, z$ and $\ket{i, \pm}$ are the eigenvectors of $\sigma_i$ with eigenvalues $\pm 1$. The action of this projector on a graph state $\ket{G}$ is given by
\begin{equation}\label{eq:2.26}
 P_{i, \pm}^{(a)} \ket{G} = \frac{1}{\sqrt{2}} \ket{i, \pm}^{(a)} \otimes U_{i, \pm}^{(a)} \ket{G'},
\end{equation}
where $G'$ is a new graph with one vertex less and $U_{i, \pm}^{(a)}$ are local unitaries \cite{Hein:2006uvf, Hein:2004zjp}. The form of the resulting graph $G'$ and the local unitaries $U_{i, \pm}^{(a)}$ vary from case to case.

\begin{figure}
 \centering
 \includegraphics[width = \linewidth]{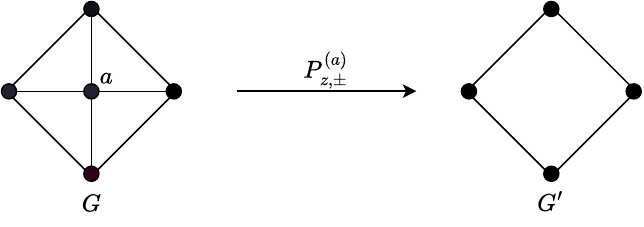}
 \caption{Acting with the projector $P_{z, \pm}^{(a)}$ at the vertex $a$ results in a graph $G'$ where the vertex $a$ is deleted.}
 \label{fig:Pz_pm}
\end{figure}

The simplest case is when $i = z$. In this case, the graph $G'$ is obtained from $G$ by deleting the vertex $a$ along with all the edges that end at it. This operation is called \emph{vertex deletion}. See figure \ref{fig:Pz_pm} for an example. The local unitaries are
\begin{equation}
 U_{z, +}^{(a)} = 1, \quad \text{and} \quad U_{z, -}^{(a)} = \prod_{b: (a, b) \in E} \sigma_z^{(b)}.
\end{equation}
In figure \ref{fig:Pz_pm}, if we act with the projection $P_{z, -}$, then we have a $\sigma_z$ sitting at each vertex of $G'$.

\begin{figure}
 \centering
 \includegraphics[width = \linewidth]{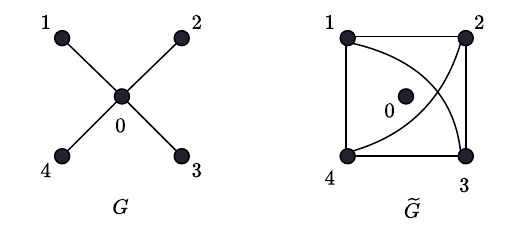}
 \caption{An example of a graph complement. The set of edges of $G$ and $\widetilde{G}$ are complementary.}
 \label{fig:graph-complement}
\end{figure}

Next we turn to the case where $i = y$. To describe the resulting graph in this case, we need some definitions.

Given a graph $G = (V, E)$, the graph complement $\widetilde{G}$ is defined as follows. For every two vertices $a$ and $b$ of $G$, if there is an edge between them, then delete it, and if there is no edge between them, then add it. The resulting graph is the graph complement $\widetilde{G}$. An example is shown in figure \ref{fig:graph-complement}.

We now define a ``local'' version of the graph complement called \emph{local complementation}. For this, we look at a vertex $a$ and consider the subgraph induced in the neighborhood of $a$. Let $N_a = \{b: (a, b) \in E\}$ be the neighborhood of $a$. Then the subgraph induced on the neighborhood of $a$ is defined as the graph $g = (N_a, E \cap (N_a \times N_a))$. Its vertices are all the neighbors of $a$, and its edges are all the edges joining neighbors of $a$ to neighbors of $a$.

\begin{figure}
 \centering
 \includegraphics[width = \linewidth]{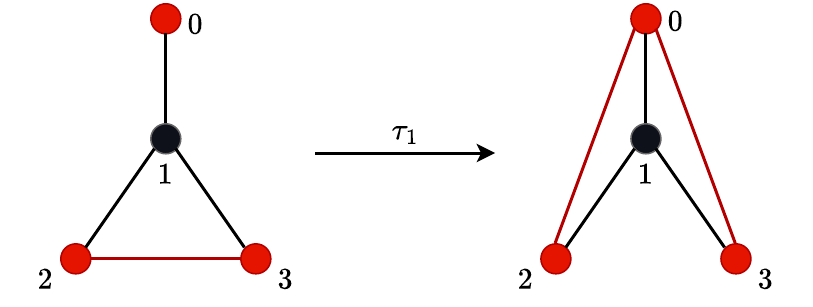}
 \caption{Local complementation at the vertex $1$. The induced subgraph in the neighborhood of $1$ is highlighted in red.}
 \label{fig:tau1}
\end{figure}

Local complementation of the graph $G$ at the vertex $a$ is obtained by complementing the subgraph induced in the neighborhood of $a$. An example is shown in figure \ref{fig:tau1}. In the left panel, the vertices $N_a = \{0, 2, 3\}$ are neighbors of $1$. These are shown in red. The only edge that connects $N_a$ to $N_a$ is $(2, 3)$ which is also shown in red. The complement of this subgraph is shown in the right panel. We deleted the edge $(2, 3)$ that was already present, and added the edges $(0, 2)$ and $(0, 3)$ which were absent. We'll denote the graph obtained via local complementation at vertex $a$ as $\tau_a(G)$.

When $i = y$, the resulting graph $G'$ is
\begin{equation}
 G' = \tau_a(G) - \{a\},
\end{equation}
i.e., local complementation at $a$ followed by deleting the vertex $a$. The local unitaries in this case are
\begin{equation}
\begin{split}
 U_{y, +}^{(a)} & = \prod_{b \in N_a} \left(-i \sigma_z^{(b)}\right)^{1/2}, \\
 \text{and} \quad U_{y, -}^{(a)} & = \prod_{b \in N_a} \left(i \sigma_z^{(b)}\right)^{1/2}.
 \end{split}
\end{equation}

When $i = x$, the situation is a little complicated. First we must pick an arbitrary vertex $b_0$ that is a neighbor of $a$. If there is no such neighbor, then $a$ is isolated and we'll deal with that case later. Having picked a $b_0$, the resulting graph is
\begin{equation}
 G' = \tau_{b_0} \circ \tau_a \circ \tau_{b_0} (G) - \{a\}.
\end{equation}
In other words, we perform a local complementation at $b_0$, followed by one at $a$, followed by one again at $b_0$, and then we delete the vertex $a$. The local unitaries are
\begin{equation}
\begin{split}
 U_{x, +}^{(a)} & = \left(i \sigma_y^{(b_0)}\right)^{1/2} \prod_{b \in N_a - N_{b_0} - \{b_0\}} \sigma_z^{(b)}, \\
 \text{and} \quad U_{x, -}^{(a)} & = \left(-i \sigma_y^{(b_0)}\right)^{1/2} \prod_{b \in N_{b_0} - N_{a} - \{a\}} \sigma_z^{(b)}.
 \end{split}
\end{equation}
When $a$ is isolated, the graph state itself is of the form $\ket{G} = \ket{+}^{(a)} \otimes \ket{G - \{a\}}$. Therefore, a $P_{x, -}^{(a)}$ collapses it to $0$, and a $P_{x, +}^{(a)}$ gives the state $\ket{G - \{a\}}$ without any local unitaries.

\subsubsection{Commutation Rules}
Our routine involves applying the projector $P_{x, +}^{(a)}$ repeatedly on the big graph $\ket{\mathcal{G}_n}$ and keeping track of the coefficient in front. From equation  (\ref{eq:2.26}), we might think that the coefficient is always $1/\sqrt{2}$, but that is not the case because of the local unitaries. Say we apply $P_{j, \pm}^{(b)}$ on equation  (\ref{eq:2.26}) where $b \neq a$. This projector must first commute through the local unitary $U_{i, \pm}^{(a)}$ before it can reach $\ket{G'}$. In \cite{Hein:2004zjp}, the authors worked out these commutation rules\footnote{At the time of writing, there are some typos in the arXiv version of \cite{Hein:2004zjp} which are corrected in the published version.} which we reproduce in Table \ref{tab:commutation}.

Owing to the non-trivial commutation relations, a projector that starts off as $P_{x, +}$ in the left can morph into another projector by the time it reaches $\ket{G'}$. Therefore, in addition to the $1/\sqrt{2}$'s that come with every projection, there are also matrix elements of the local unitaries that we need to keep track of.

\begin{table*}
\centering
\begin{tabular}{|c| c| c|}
\hline
$P_{x,\pm} \sigma_z = \sigma_z P_{x,\mp}$ & $P_{y, \pm} \sigma_z = \sigma_z P_{y, \mp}$ &
$P_{z,\pm} \sigma_z = \sigma_z P_{z, \pm}$ \\
\hline
$P_{x,\pm} (-i \sigma_z)^{1/2} = (-i \sigma_z)^{1/2} P_{y, \mp}$ & $P_{y, \pm} (-i \sigma_z)^{1/2} = (-i \sigma_z)^{1/2} P_{x,\pm}$ & $P_{z, \pm} (-i \sigma_z)^{1/2} = (-i\sigma_z)^{1/2} P_{z,\pm}$ \\
\hline
$P_{x,\pm} (i \sigma_y)^{1/2} = (i \sigma_y)^{1/2} P_{z, \mp}$ & $P_{y, \pm} (i \sigma_y)^{1/2} = (i\sigma_y)^{1/2} P_{y, \pm}$ & $P_{z,\pm} (i\sigma_y)^{1/2} = (i \sigma_y)^{1/2} P_{x, \pm}$ \\
\hline
$P_{x,\pm} (-i \sigma_y)^{1/2} = (-i \sigma_y)^{1/2} P_{z,\pm}$ & $P_{y,\pm} (-i \sigma_y)^{1/2} = (-i \sigma_y)^{1/2} P_{y,\pm}$ & $P_{z,\pm} (-i \sigma_y)^{1/2} = (-i \sigma_y)^{1/2} P_{x,\mp}$ \\
\hline
$P_{x,\pm} (i \sigma_z)^{1/2} = (i \sigma_z)^{1/2} P_{y, \pm}$ & $P_{y, \pm} (i \sigma_z)^{1/2} = (i \sigma_z)^{1/2} P_{x,\mp}$ & $P_{z,\pm} (i \sigma_z)^{1/2} = (i \sigma_z)^{1/2} P_{z,\pm}$\\
\hline
\end{tabular}
\caption{Commutation rules for the projectors and local unitaries. Reproduced from \cite{Hein:2004zjp}.}
\label{tab:commutation}
\end{table*}

\subsubsection{Summary of the Numerical Routine}
\begin{enumerate}
 \item Given a graph state $\ket{G}$, an integer $n$, and a partition of its qubits into $q$ parties, we first construct the big graph $\mathcal{G}_n$. The multi-invariant $\mathcal{Z}$ of the graph state $\ket{G}$ is the inner product in equation  \eqref{eq:inner-product}.

 \item To compute the inner product, we apply projectors $\ket{+}\bra{+}$ iteratively for each vertex of $\mathcal{G}_n$. At each step, we get a reduced graph $G'$ and some local unitaries.

 \item We commute successive projectors through the unitaries according to the rules in Table \ref{tab:commutation} and keep track of the coefficient in the front.

 \item In the end, we're left with $\mathcal{Z} \ket{+ \dots +}$. The final coefficient  is the inner product.
\end{enumerate}

\begin{figure}
    \centering
    \includegraphics[width=\linewidth]{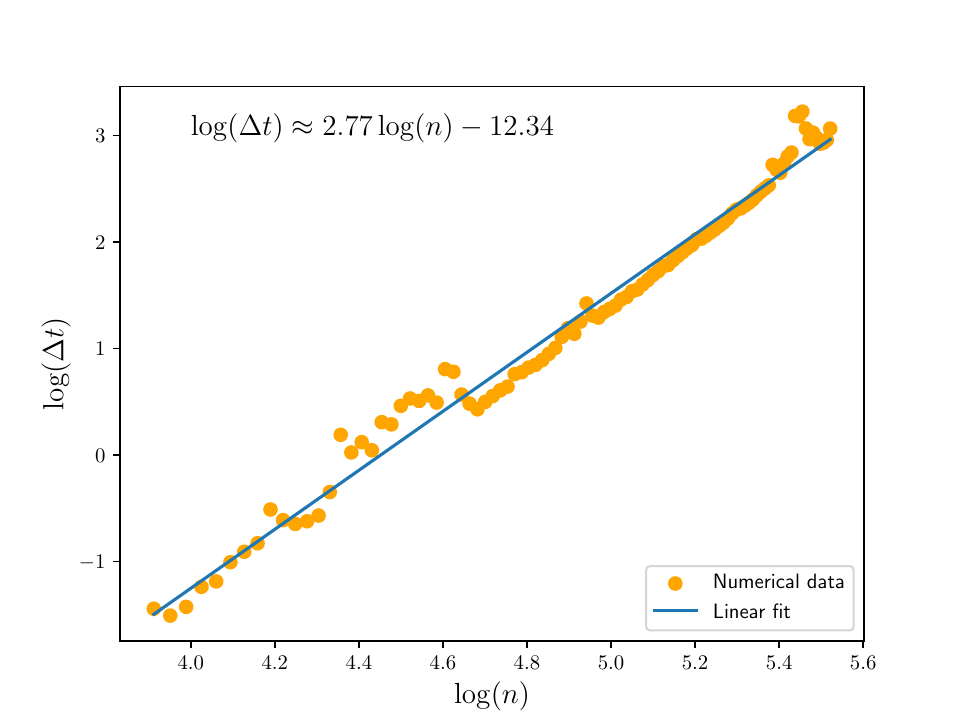}
    \includegraphics[width=\linewidth]{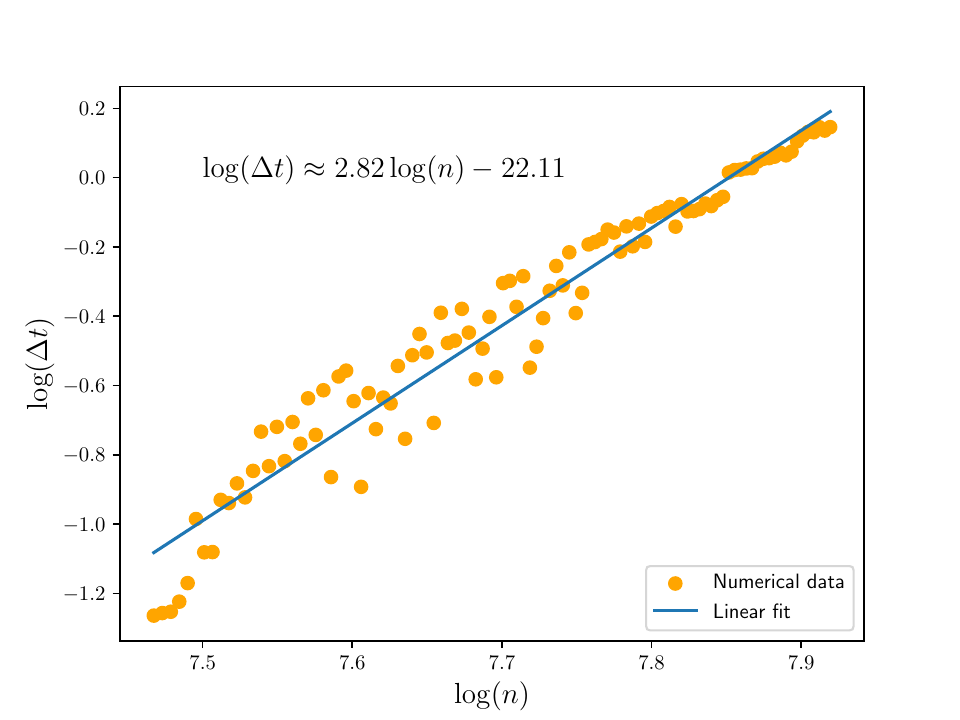}
    \caption{Time complexity of our algorithm and the algorithm in \cite{garcia2013efficientinnerproductalgorithmstabilizer} respectively. Here $n$ is the number of vertices in the graph.}
    \label{fig:time-complexity-python}
\end{figure}

After we came up with our algorithm, we came across another efficient inner product algorithm in \cite{garcia2013efficientinnerproductalgorithmstabilizer}. We review their approach in appendix \ref{another}. Comparison of two algorithms is shown in figure  \ref{fig:time-complexity-python}. Both algorithms go roughly like $\mathcal{O}(n^3)$ where $n$ is the number of vertices in the graph.

\section{Tripartite Stabilizer States}\label{sec:tripartite-stabilizer-states}
In this section, we show that tripartite stabilizer states admit simple analytic expressions for their multi-invariants \cite{Akella:2025owv}. This works due to the following theorem in \cite{Bravyi_2006}:

\begin{theorem}\label{ghz-extraction}
    Any tripartite stabilizer state is LC-equivalent to a collection of: (a) GHZ states, (b) Bell pairs, and (c) unentangled states.
\end{theorem}

We can even quantify the number of GHZs, Bell pairs, and unentangled states that can be extracted from $\ket{\psi}$. To do this, let's define the following subgroups of the full stabilizer group $G$ of $\ket{\psi}$. The qubits of $\ket{\psi}$ are partitioned into three parties $A$, $B$, and $C$. Let $G_A$ be the subgroup of $G$ consisting of all elements that have a trivial action on the qubits in $B$ and $C$. Similarly, let $G_{BC}$ be the subgroup of all elements which have a trivial action on $A$ and so on. We can then define the following subgroup:
\begin{equation}
    G_{AB} \cdot G_{BC} = \{ gh : g \in G_{AB}, \, h \in G_{BC}\}.
\end{equation}
With these definitions, the number of GHZs that can be extracted from a tripartite stabilizer state $\ket{\psi}$ is
\begin{equation}
    p = \log_2 \left(\frac{|G|}{|G_{AB} \cdot G_{BC} \cdot G_{AC}|}\right).
\end{equation}
Similarly, the number of Bell pairs that can be extracted between parties $A$ and $B$ is
\begin{equation}
    m_{AB} = \frac{1}{2}\left(\log_2\left(\frac{|G_{AB}|}{|G_A||G_B|}\right) - p\right),
\end{equation}
and similarly for $m_{BC}$ and $m_{AC}$.

This theorem allows us to write any multi-invariant of a tripartite stabilizer state as
\begin{align}
    &{\cal Z}(\sigma_1,\sigma_2,\sigma_3)=({\cal Z}_{\rm GHZ}(\sigma_1,\sigma_2,\sigma_3))^{p}\\
    &({\cal Z}_{\rm Bell}(\sigma_1,\sigma_2))^{m_{AB}}({\cal Z}_{\rm Bell}(\sigma_2,\sigma_3))^{m_{BC}}({\cal Z}_{\rm Bell}(\sigma_1,\sigma_3))^{m_{AC}}\notag
\end{align}
Here ${\cal Z}_{\rm GHZ}(\sigma_1,\sigma_2,\sigma_3)$ stands for the multi-invariant ${\cal Z}(\sigma_1,\sigma_2,\sigma_3)$ evaluated on the GHZ state, ${\cal Z}_{\rm Bell}(\sigma_1,\sigma_2)$ stands for the bipartite multi-invariant ${\cal Z}(\sigma_1,\sigma_2)$ evaluated on the Bell state.

As a result, the R\'enyi multi-entropy for a tripartite stabilizer state is given by the following formula
\begin{align}
    \mathcal{E}_n^{\text{ME}} &= \log_2 \Big(\frac{|G|}{(|G_A| |G_B| |G_C|)^{n/2}}\notag\\
     &\times |G_{AB} \cdot G_{BC} \cdot G_{AC}|^{(n-2)/2}\Big).
\end{align}
This admits a straightforward analytic continuation to $n=1$ that gives us the multi-entropy.

An interesting feature of the above formula is that when $n = 2$, the cardinality of the subgroup $G_{AB}\cdot G_{BC} \cdot G_{AC}$ drops out of the expression. In this case, the above expression can be derived via a simple counting argument given below. The underlying multi-invariant in $n=2$ case is a type of Coxeter multi-invariant. These invariants are reviewed in section \ref{subsec:coxeter}.
We conjecture a simple closed form expression for the Coxeter invariant of any, even higher-partite, stabilizer state in section \ref{subsec:conjecture}.

\subsection{Counting argument for the second R\'enyi multi-entropy}\label{counting}
To compute any multi-invariant for a stabilizer state, it is convenient to use the expression
\begin{align}
    \ket{\psi}\bra{\psi} = \frac{1}{|G|} \sum_{\sigma \in G} \sigma,
\end{align}
for the density matrix rather than using the state itself. Consequently, it is useful to express the multi-invariant as a function of this density matrix rather than the state. For this we use the doubling trick introduced before. Instead of computing the desired multi-invariant ${\cal Z}$, we compute $|{\cal Z}|^2$. Graphically, $|{\cal Z}|^2$ is obtained by superimposing the graph of ${\cal Z}$ with that of the complex conjugate graph (vertex parity exchanged). The resulting graph now has the density matrix $|\psi\rangle\langle \psi|$ ``living at'' the vertex. This doubling trick is illustrated for the case of bipartite second R\'enyi entropy in figure \ref{fig:Renyi-doubling}. In all the cases we consider ${\cal Z}$ is real and positive, so we can replace $|{\cal Z}|^2$ by ${\cal Z}^2$.

In this section, we derive the formula for the $n = 2$ R\'enyi multi-entropy, i.e., $\mathcal{E}_2^{\text{ME}}$, of a tripartite stabilizer state via a simple counting argument. As a warm-up, let us consider the second R\'enyi entropy of a bipartite stabilizer state.

\subsubsection*{Warm-up: Bipartite}
\begin{figure}	\includegraphics[width = 0.9 \linewidth]{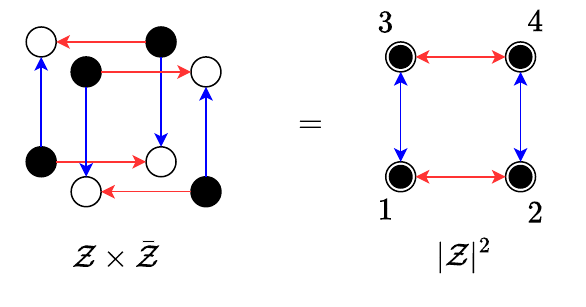}
\caption{The doubling trick applied to the $n = 2$ R\'enyi entropy. The rank-one projector $\ketbra{\psi}$ lives at each vertex of the doubled graph.}
\label{fig:Renyi-doubling}
\end{figure}

The square of ${\rm Tr} \rho_A^2$  is denoted graphically in figure \ref{fig:Renyi-doubling}. At each vertex of the square, we have a copy of a stabilizer state
\begin{equation}
	\ketbra{\psi} =  \frac{1}{|G|} \sum_{\sigma \in G} \sigma.
\end{equation}
Each term in the above sum can be factorized as $\sigma=\sigma_A\otimes \sigma_B$ where $\sigma_A$ are the Pauli operators supported over the qubits in party $A$, and so on. Let us understand the form of $\sigma_A$ and $\sigma_B$ for a graph state. The stabilizer group of a graph state is generated by elements of the form
\begin{equation}
	g_a = X_a \prod_{b} Z_b^{\gamma_{ab}}
\end{equation}
where $\gamma$ is the adjacency matrix of the graph.
Each element can then be labeled by a binary vector $\vec u$ of length $n$ that indicates whether $g_a$ at site $a$ is ``turned on''. More explicitly, $g(\vec u):=\prod_{a:u_a=1}g_a$. Up to minus sign, this element takes the form $g(\vec u)=X^{\vec u}Z^{\gamma\cdot \vec u}$, where we have used the notation $X^{\vec u}= X_1^{u_1}\otimes \ldots X_n^{u_n}$, and similarly for $Z$. If we break up $\vec u$ as $(u_A,u_B)$ where $u_A$  is the part of $\vec u$ supported over qubits in $A$ and so on, then we have, up to minus signs,
\begin{align}\label{sigma_explicit}
    \sigma_A=X^{u_A}Z^{(\gamma \cdot \vec u)_A},\quad \sigma_B=X^{u_B}Z^{(\gamma \cdot \vec u)_B}.
\end{align}
Now we are in position to compute the R\'enyi invariant ${\cal Z}^2$ given in figure \ref{fig:Renyi-doubling}.
\begin{align}
    {\cal Z}^2&= {\rm Tr}(\sigma_{1A}\sigma_{2A}){\rm Tr}(\sigma_{1B}\sigma_{3B})\notag\\
    &\times {\rm Tr}(\sigma_{3A}\sigma_{4A}){\rm Tr}(\sigma_{2B}\sigma_{4B}).
\end{align}
We have used the notation $\sigma_1=\sigma_{1A}\otimes \sigma_{1B}$ etc..
In each term ${\rm Tr}(\sigma \sigma')$ is non-zero if and only if $\sigma=\sigma'$. As $\sigma'$ has to be the same as $\sigma$, the overall minus sign of $\sigma$ gets cancelled in the trace; so we don't need to keep track of it and we can simply use the form of the $\sigma$'s computed in equation \eqref{sigma_explicit}. Let us solve the constraints coming from each of the four traces.

First consider the constraint $\sigma_{1A}=\sigma_{2A}$. Taking ${\vec u}_1=(u_A,u_B)$, we see that ${\vec u}_2$ has to be $(u_A, u_B+r_B)$ where $r_B$ obeys the condition
\begin{align}
    \gamma_{AB}\cdot r_B=0.
\end{align}
where $\gamma_{AB}$ is defined as follows. Split the adjacency matrix $\gamma$ of the graph state corresponding to $\ket{\psi}$ into the following block diagonal form:
\begin{equation}
	\gamma = \begin{pmatrix}
		\gamma_{AA} & \gamma_{AB} \\ \gamma_{BA} & \gamma_{BB}
	\end{pmatrix}.
\end{equation}
The number of solutions for $r_B$ is precisely $|G_B|$. This is easy to see from equation \eqref{sigma_explicit}. Note that $\sigma_A=1$ implies $u_A=0$ and $\gamma_{AB}\cdot u_B=0$. This shows each element of $G_B$ is parametrized by that $\vec u= (0,r_B)$ with $\gamma_{AB}\cdot r_B=0$.
Next, consider the constraint $\sigma_{1B}=\sigma_{3B}$. In the same way as before, this solution is solved by ${\vec u}_3=(u_A+s_A, u_B)$ where $s_A$ obeys $\gamma_{BA}\cdot s_A=0$. The number of solutions to this equation is $|G_A|$. After solving for $\vec u_3$ and $\vec u_4$, in terms of $\vec u_1$, $\vec u_4$ is uniquely determined to be $(u_A+s_A, u_B+r_B)$. Collecting all the solutions, figure \ref{fig:Renyi-doubling} evaluates to
\begin{equation}\label{bipartite-counting}
	\mathcal{Z}^2 = \frac{1}{|G|^4} \times |G||G_A||G_B| \times 2^{2|A|} 2^{2|B|}.
\end{equation}
The factor of $|G|^4$ in the denominator comes from the normalization of each of the four density matrices. The factors of $2^{|A|}$ and $2^{|B|}$ come from the traces of the identity over $A$ and $B$ respectively. From our convention for the multi-entropy in equation  (\ref{eq:Renyi-multis}), we get
\begin{equation}
	\mathcal{E}_{n = 2, q = 2}^{\text{ME}} = \frac{1}{2} \log \left(\frac{|G|}{|G_A| |G_B|} \right)
\end{equation}
which in fact matches the entanglement entropy derived in equation  (\ref{eq:ent-ent}).

This counting argument can also be made for general stabilizer states that are not necessarily in graph state form. In fact, in its general form, the logical structure of the argument becomes more transparent. We first turn on a general element $g$ of the stabilizer group at vertex $1$. Non-vanishing contribution is obtained by taking the stabilizer element to be the same at all vertices. Fixing the stabilizer element to be $g$ at $1$, we ask what other freedom do we have to turn on stabilizer elements. We can turn on an element $g'$ at vertex $2$. Because it should not change the element at $1$, $g'$ must only be supported on $B$ i.e. $g'\in G_{B}$. To get non-vanishing contribution, the same $g'$ needs to be turned on at vertex $4$. Now we look for more freedom. We can turn on $g''$ at vertex $3$. Because it should not affect $g$ at $1$, $g''\in G_A$. This also forces an extra factor of $g''$ at vertex $4$. There is no further freedom to turn on any stabilizer elements at vertex $4$ without affecting the choices at vertices $1,2$ and $3$. This argument precisely reproduces the formula \eqref{bipartite-counting}.

The first assignment of $g\in G_{AB}$ is to all the vertices of the graph. We associate it to the 2-cell of the graph.
The assignment of $g'\in G_B$ is to the $B$-colored subgraph with vertices $2$ and $4$ while the assignment of $g''\in G_A$ is to the $A$-colored subgraph with vertices $3$ and $4$. Both of these are 1-cells of the graph.
The counting argument for tripartite state also follows extension of this assignment scheme as we will see shortly.

\subsubsection*{Tripartite}

\begin{figure}[b]
	\includegraphics[width = 0.4\linewidth]{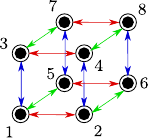}
	\caption{The grid that calculates the $n = 2$, $q=3$ R\'enyi multi-entropy using the doubling trick. The rank-one projector $\ketbra{\psi}$ lives at each vertex.}
	\label{fig:cube-grid}
\end{figure}

We compute the R\'enyi multi-entropy ${\cal E}_2$ for the tripartite state using the general argument that we distilled towards the end of the previous subsection. A more refined argument, specialized to graph states, is given in appendix \ref{graph_state_counting}.

The tripartite stabilizer state is given as,
\begin{equation}
	\ket{\psi}\bra{\psi} = \frac{1}{|G|} \sum_{\sigma \in G} \sigma,
\end{equation}
where $G$ is the stabilizer group of $\ket{\psi}$. As before, the stabilizer subgroup that has support only over $A$ is denoted as $G_A$ etc. and that has support over both $A$ and $B$ is denoted as $G_{AB}$ etc.

We turn on a general element $g$ at vertex $1$. The constraint is satisfied by having the same element at every vertex. While looking for additional freedom, we make sure that the choice at vertex $1$ is kept fixed to avoid over-counting. We turn on $g'\in G_{BC}$ at vertex $2$, the constraint is solved by having the same element at vertex $2,4,6$ and $8$. Similarly we turn on elements in $G_{AB}$ and $G_{AC}$ at vertices $5,6,7,8$ and $3,4,7,8$ respectively. Now, as we look for additional freedom, we require that the freedom at vertex $2$ must be identity in $G_{BC}$ to avoid over-counting and similarly, the freedom at vertex $5$ and $3$ must be identity in $G_{AB}$ and $G_{AC}$ respectively. This constraint can be solved by turning on an element of  $G_B$ at vertex $6$ and $8$,  element of $G_A$ at vertices $7$ and $8$ and element of $G_C$ at vertices $4$ and $8$. After this assignment, there is no freedom left at $8$. All in all we get,
\begin{align}\label{tripartite-counting}
    {\cal Z}^2=\frac{1}{|G|^3}|G_{AB}||G_{BC}||G_{AC}||G_A||G_B||G_C|.
\end{align}
If we use the following identity that these cardinalities satisfy:
\begin{equation}
	|G_{AB}||G_{BC}| |G_{AC}| = |G||G_A||G_B||G_C|,
\end{equation}
we get,
\begin{equation}
	\mathcal{E}_{n = 2, q = 3}^{\text{ME}} =  \log \left(\frac{|G|}{|G_A| |G_B| |G_C|} \right)
\end{equation}

The first assignment of $g\in G_{ABC}$ is to all the vertices of the graph. We associate it to 3-cell of the graph. The assignment of $g'\in G_{BC}$ is to the $BC$-colored subgraph with vertices $2,4,6,8$.
The assignment of elements of $G_{AB}$ is to the $AB$-colored subgraph with vertices $5,6,7,8$. The assignment of elements of $G_{AC}$ is to the $AC$-colored subgraph with vertices $3,4,7,8$. All these are 2-cells of the graph.
The assignment of $g''\in G_A$ is to the $A$-colored subgraph with vertices $7$ and $8$, and so on. All these are 1-cells of the graph.

We conjecture that this assignment scheme generalizes to the so-called Coxeter invariants. Before we state the conjecture, we will review the Coxeter invariants ${\cal Z}_{\rm cox}$. It has been shown in \cite{Gadde:2025csh}, $1-{\hat {\cal Z}}_{\rm cox}$ is an entanglement monotone. Here $\hat {\cal Z}$ is defined as ${\cal Z}^{1/n}$ where $n$ is the number of density matrices in ${\cal Z}$.

\subsection{Coxeter groups and Coxeter invariants}\label{subsec:coxeter}
A Coxeter group is a group generated by a set of generators $\{r_1, \dots, r_n\}$ that satisfy:
\begin{equation}
(r_i r_j)^{m_{ij}} = 1
\end{equation}
where $m_{ii} = 1$ and $m_{ij} \geq 2$ for $i \neq j$. These groups are a generalization of reflection groups \cite{coxeter1934discrete}. In fact, finite Coxeter groups are reflection groups and Coxeter classified all of them in \cite{coxeter1935complete}. However, the above definition also admits infinite groups. A quick way to see this is by setting $m_{ij} = \infty$, in which case the element $r_i r_j$ has infinite order.

The matrix $(m_{ij})$ is a symmetric $n \times n$ matrix with $1$'s along the diagonal and is called the Coxeter matrix. It can be viewed as the adjacency matrix of a \emph{weighted graph}.

A weighted graph is a graph $(V, E)$ with an associated weight function $w: E \to \mathbb{R}$ that assigns a weight $w(e)$ to each edge $e \in E$. In our case, the adjacency matrix $m_{ij}$ defines a graph with $n$ vertices with the edge between $i$ and $j$ carrying the weight $w(i, j) = m_{ij}$.

\begin{figure}
	\centering
	\includegraphics[width = \linewidth]{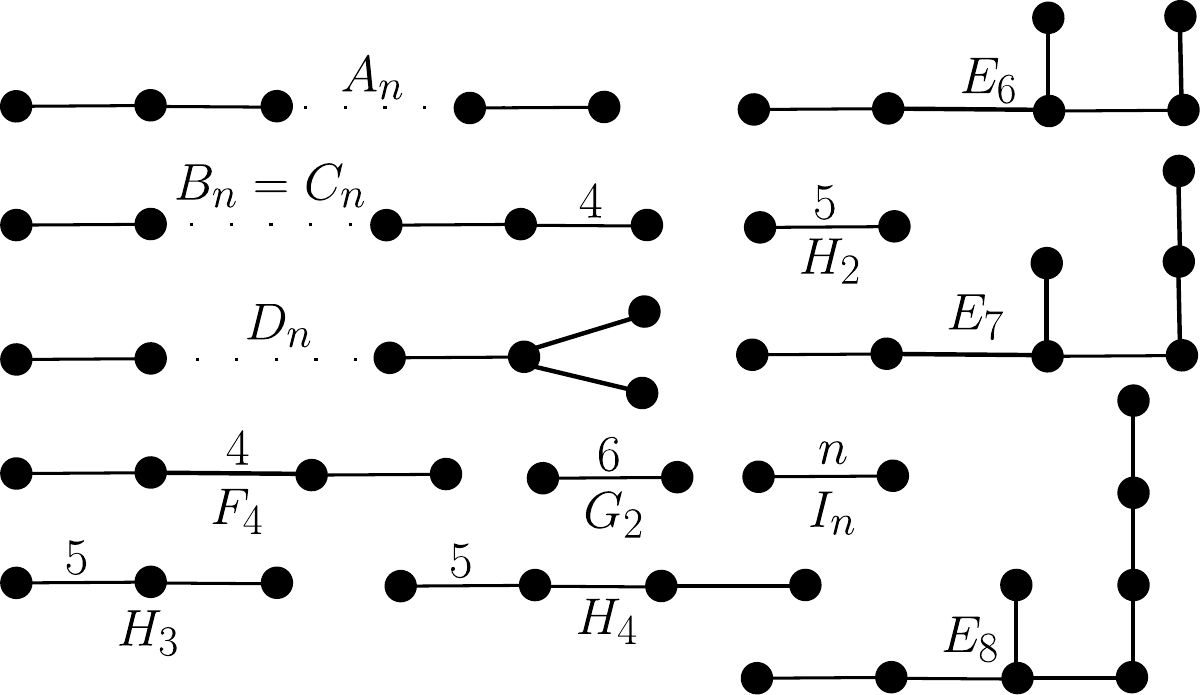}
	\caption{The Coxeter-Dynkin diagrams of all irreducible finite Coxeter groups.}
	\label{fig:finite-coxeter}
\end{figure}

We make the following simplifications to this graph:
\begin{itemize}
	\item Delete the edges with weight $2$. When there is no edge between vertices $i$ and $j$, we mean that $m_{ij} = 2$.
	\item Don't label the edges with weight $3$. When there is an edge between vertices $i$ and $j$ with no label, we mean that $m_{ij} = 3$.
\end{itemize}
The other edges are labelled by their weight which is at least $4$. This defines the \emph{Coxeter-Dynkin diagram} corresponding to a Coxeter group. All finite Coxeter groups have been classified by their Dynkin diagrams, and they are the disconnected sums of the Dynkin diagrams shown in figure \ref{fig:finite-coxeter}.

A generalization of the $n = 2$ R\'enyi multi-entropy would be a multi-invariant of the form $\mathcal{E}(r_1, \dots, r_q)$ where $\{r_1, \dots, r_q\}$ are picked from the generators of a finite Coxeter group. However, the group elements $r_i$ don't look like simple permutations. To turn them into permutations, we need to define a set on which they can act like permutations, and the \emph{Cayley graph} \cite{cayley1878desiredata} does just that.

Define a finite group $\mathcal{K}$ as the one generated by the elements $r_1, \dots, r_q$. Associate a vertex to each element of $\mathcal{K}$ and add an edge between two vertices $g_1$ and $g_2$ if they are related via $g_2 = r_i g_1$ for some generator $r_i$. The action of any group element $g \in \mathcal{K}$ now acts like a permutation on the vertices of the Cayley graph of $\mathcal{K}$, and this action is free and transitive. The coordination number of each vertex is $q$, and the Cayley graph is starting to look a lot like the bipartite graphs we were using to define multi-invariants.

The Cayley graph of $\mathcal{K}$ is bipartite. To see this, define the length of any group element $g \in \mathcal{K}$ as the distance on the Cayley graph between $g$ and the identity. The vertices of the Cayley graph fall into two categories: those with even length, and those with odd length. We imagine placing $q$-partite states $\psi_{\alpha_1 \dots \alpha_q}$ on the even vertices, and $\bar{\psi}^{\beta_1 \dots \beta_q}$ at the odd sites. This therefore defines the multi-invariant $\mathcal{Z}_{\rm cox}(r_1, \dots, r_q)$. The multi-invariant for $q$-partite $n=2$ R\'enyi multi-entropy corresponds to the Dynkin diagram that is disconnected with $q$ nodes and has no edges.

\subsection{Conjecture}\label{subsec:conjecture}

We conjecture that ${\cal Z}_{\rm cox}$ is computed using  the same assignment scheme that we used for $n=2$ R\'enyi multi-entropy for bipartite and tripartite states. We will associate $|G|$ to the full graph. Then we will associate $G_{\hat A}$ to all but one of the $q-1$ colored subgraphs ${\cal Z}_{\hat A}$ where $\hat A$ stands for all the color labels except for $A$. This gives rise to the factor $|G_{\hat A}|^{n_{\hat A}-1}$ where $n_{\hat A}$ is the number of $q-1$ colored subgraphs ${\cal Z}_{\hat A}$ with colors in ${\hat A}$. The power is $n_{\hat A}-1$ and not $n_{\hat A}$ because, the assignment of an element of $G_{\hat A}$ to all $q-1$ subgraphs could be absorbed into the initial assignment of $g$ to all the vertices. Similarly, we assign an element of $G_{\hat B}$ to all $q-1$ colored subgraphs ${\cal Z}_{\hat B}$ etc. Then we move on to $q-2$ colored subgraphs. To each such subgraph, say ${\cal Z}_{\hat A\hat B}$, we assign $g\in G_{\hat A\hat B}$. This gives rise to over-counting as before because some of these factors can be absorbed into either the assignment of $g\in G_{\hat A}$ or the assignment of $g\in G_{\hat B}$. This over-counting removes the power $n_{\hat A}+n_{\hat B}$ of the factor $|G_{\hat A\hat B}|$. But now we have overcompensated because one such factor was not even counted in the assignment of $G_{\hat A}$ and $G_{\hat B}$. All in all, we get the power $G_{\hat A\hat B}^{n_{\hat A\hat B}-n_{\hat A}-n_{\hat B}+1}$. We compute powers of all the stabilizer subgroups using such inclusion exclusion principle.
To state the result, it is useful to introduce the following quantity that is associated to a general $q$-partite stabilizer state specified by the group $G=G_{ABC\ldots}$,
\begin{align}
    k_{G}=\frac{|G|}{(|G_{\hat A}||G_{\hat B}|\ldots)}\times \frac{(|G_{\hat A\hat B}||G_{\hat A\hat C}|\ldots) }{(|G_{\hat A\hat B\hat C}|| G_{\hat A\hat B\hat D}|\ldots) }\times \ldots
\end{align}
Here each bracketed factor is a product over all combinations of the subscripts. The result of this counting is that total number of assignments for a general Coxeter graph is
\begin{align}
    k_G\cdot ( k_{G_{\hat A}}^{n_{\hat A}}\cdot  k_{G_{\hat B}}^{n_{\hat B}}\ldots )\cdot(k_{G_{\hat A\hat B}}^{n_{\hat A\hat B}}\cdot k_{G_{\hat A\hat C}}^{n_{\hat A\hat C}}\ldots)\ldots
\end{align}
Here also each bracketed factor is a product over all combination of the subscripts. We divide this quantity by the normalization  $|G|^n$ where $n$ is the total number of replicas in the Coxeter graph to get multi-invariant squared ${\cal Z}_{\rm cox}^2$.

In the case of the $n=2$ R\'enyi multi-invariant for $q=2$, we have $n_{\hat A}=n_{\hat B}=2$. So we have
\begin{align}
    {\cal Z}_{\rm cox}^2=\frac{1}{|G|^2} \frac{|G|}{|G_{A}||G_{B}|} |G_{A}|^{2}|G_B|^2=\frac{|G_A||G_B|}{|G|}.
\end{align}
For $q=3$, we have $n_{\hat A}=n_{\hat B}=n_{\hat C}=2$ and $n_{\hat A\hat B}=n_{\hat A\hat C}=n_{\hat B\hat C}=4$. We have
\begin{align}
    {\cal Z}_{\rm cox}^2&=\frac{1}{|G|^4}\frac{|G||G_{A}||G_{B}||G_{C}|}{|G_{AB}||G_{AC}||G_{BC}|}\times \Big(\frac{|G_{AB}|}{|G_{A}||G_B|}\ldots\Big)^2\notag\\
    &\times (|G_A||G_B||G_C|)^4\notag\\
    &= \frac{|G_A||G_B||G_C||G_{AB}||G_{BC}||G_{CA}|}{|G|^3}.
\end{align}
This is exactly the same answer obtained by explicit counting in equation \eqref{tripartite-counting}. Interestingly, the form of ${\cal Z}^2_{\rm cox}$ admits a rewriting as a product over all bipartitions. Denoting a subset  of $\{A,B,C\}$ as $R$,
\begin{align}\label{eq:bipartitions}
    {\cal Z}_{\rm cox}^2=\prod_R \frac{|G_R||G_{\hat{R}}|}{|G|}.
\end{align}
Here $\hat R$ is the complement of $R$. In this product, we count each pair $(R,\hat R)$ once. The higher partite $n=2$ R\'enyi multi-invariants also can be massaged into this form. In that case, $R$ is taken to be a subset of the set of $q$-parties and $\hat R$, its complement. As a result, $n=2$ R\'enyi multi-entropy is a sum over entanglement entropies across all bipartitions.
It would be interesting to understand the reason for this simplification.

\subsubsection{Other tripartite Coxeter invariants}
If we restrict ourselves to three parties, i.e., $q = 3$, and pick generators $r_A, r_B$ and $r_C$ from a Coxeter group, we get a tripartite Coxeter invariant $\mathcal{Z}(r_A, r_B, r_C)$. From the classification of finite Coxeter groups in figure \ref{fig:finite-coxeter}, we see that only the following groups are allowed: $A_3$, $B_3$, $H_3$, and the group with a disconnected Coxeter-Dynkin diagram made up of $I_n$ and an isolated vertex. The orders of the elements $r_A r_B$ in each of these groups is captured by the triples $(m_{AB},m_{BC},m_{CA})$= $(2, 3, 3)$, $(2, 3, 4)$, $(2, 3, 5)$, and $(2, 2, n)$ respectively.

Let us take the example of the case $(2,3,3)$. Here $n_{AB}=6,n_{BC}=4,n_{CA}=4$ and $n_A=n_B=n_C=12$. Substituting in our formula, we get
\begin{align}
    {\cal Z}_{{\rm cox},(2,3,3)}^2=\left(\frac{|G_A|^7 |G_B|^7 |G_C|^8}{|G_{AB}|^3 |G_{BC}|^4 |G_{AC}|^4} \right)^2
\end{align}
Similarly, for other cases we have
\begin{align}
    {\cal Z}_{{\rm cox},(2,3,4)}^2&=\left(\frac{|G_A|^{15} |G_B|^{14} |G_C|^{17}}{|G_{AB}|^6 |G_{BC}|^8 |G_{AC}|^9} \right)^2\notag\\
    {\cal Z}_{{\rm cox},(2,3,5)}^2&=\left(\frac{|G_A|^{39} |G_B|^{35} |G_C|^{44}}{|G_{AB}|^{15} |G_{BC}|^{20} |G_{AC}|^{24}}\right)^2\notag\\
    {\cal Z}_{{\rm cox},(2,2,n)}^2&=\left(\frac{|G_A|^n |G_B|^{3n/2 - 1} |G_C|^{3n/2 - 1}}{|G_{AB}|^{n/2} |G_{BC}|^{n-1} |G_{AC}|^{n/2}} \right)^2\notag.
\end{align}

Note that none of these expressions contain the factor $|G_{AB}\cdot G_{BC}\cdot G_{AC}|$. This factor certainly appears in general tripartite multi-invariant. In fact the presence of this factor can also be seen in R\'enyi multi-entropy for $n\geq 3$. These multi-invariants are not of Coxeter type.

It turns out that the factor $|G_{AB}\cdot G_{BC}\cdot G_{AC}|$ captures the topology of the multi-invariant graph. We will show this below. In order to make connection with topology, we associate a $2$-manifold with any $3$-partite multi-invariant. This is done by taking the tripartite state to be a state on a triangle with each party being one side of the triangle. The edge contraction pattern of the multi-invariant gives a gluing pattern for the triangles. As a result, multi-invariant gives rise to a bicolorable triangulation of a $2$-manifold.

We will show that the factor $|G_{AB}\cdot G_{BC}\cdot G_{AC}|$ appears with power ${\mathtt g}$ where ${\mathtt g}$ is the genus of this surface. We will do so by computing a general multi-invariant ${\cal Z}={\cal Z}(\sigma_1,\sigma_2,\sigma_3)$ for GHZ states and Bell pairs and using theorem \ref{ghz-extraction} to express it in terms of order of stabilizer subgroups. For the GHZ state
\begin{align}
    \frac{1}{\sqrt 2}(|000\rangle+|111\rangle),
\end{align}
all vertices of the multi-invariant graph are either ``occupied'' by all $0$s or all $1$s. Correspondingly we get the contribution,
\begin{align}
    {\cal Z}_{\rm GHZ}=2^{1-N/2}
\end{align}
where $N$ is the total number of vertices of the multi-invariant. Now let us consider the Bell-pair between the parties $A$ and $B$. The contribution can be computed by erasing the $C$ edges and obtaining a $n_{AB}$ number of $AB$-colored subgraphs \footnote{We will assume that all these subgraphs are isomorphic for simplicity, the extension to more general case is straightforward.}. The contribution is then
\begin{align}
    {\cal Z}_{{\rm Bell}_{AB}}=2^{n_{AB}-N/2}.
\end{align}
Using theorem \ref{ghz-extraction}, we now write the general form of the multi-invariant in terms of $n_{AB}, n_{BC}, n_{CA}$. Note that $n_A=n_B=n_C=N/2$.
\begin{align}
    {\cal Z}&=\frac{(|G_A|^{\frac12(N-n_{AB}-n_{AC})}\ldots)}{(|G_{AB}|^{\frac12(\frac{N}{2}-n_{AB})}\ldots)}\notag\\
    &\times |G_{AB}\cdot G_{BC}\cdot G_{AC}|^{\frac12(n_{AB}+n_{BC}+n_{CA}-\frac{N}{2}-2)}
\end{align}
Using the formula for the Euler characteristic,
\begin{align}
    \chi&=V-E+F\notag\\
    &=N-\frac{3N}{2}+n_{AB}+n_{BC}+n_{CA}\notag\\
    &=2-2{\mathtt g}
\end{align}
we see that the power of $|G_{AB}\cdot G_{BC}\cdot G_{AC}|$ is ${\mathtt g}$. We expect that our conjecture for general Coxeter group holds because the resulting multi-invariants describe higher dimensional manifolds with trivial topology.

\section{Coxeter multi-invariants of X-stabilizer states}\label{sec:restricted}
In this section, we consider a restricted class of stabilizer states where the counting argument from before gives a much simpler result. In the previous section we saw how the cardinalities of the various subgroups of the full stabilizer group appear when calculating multi-invariants. However, the stabilizer group for an $n$-qubit state has an order of $2^n$ which is too large for numerics. Within the class of stabilizer states, there is a special class of states which are described by an even smaller abelian group $N \subset G$ that completely describe the state. We refer to these states as \emph{X-stabilizer states}. 

An X-stabilizer state is defined as follows. Let $N$ be a subgroup of the $n$-qubit Pauli group $\mathcal{P}_n$ made of Pauli strings containing only $I$ and $X$: the identity and the Pauli $X$ operator, with no signs. Define the state
\begin{equation}
    \ket{\mathcal{N}} = \frac{1}{|N|^{1/2}} \sum_{g \in N} g \ket{0}^{\otimes n},
\end{equation}
where $\ket{0}$ is the eigenstate of the Pauli operator $Z$ with eigenvalue $+1$. Acting with $X$ flips it: $X\ket{0}=\ket{1}$. Since $N$ is a group, any element of $N$ stabilizes $\ket{\mathcal{N}}$. We can easily show that $\ket{\mathcal{N}}$ is a stabilizer state by writing down the full stabilizer group of order $2^n$. Let
\begin{equation}
    X^{\vec{u}} = X^{u_1} \otimes \dots \otimes X^{u_n}
\end{equation}
belong to $N$. The set of all such $\vec{u}$'s that define $N$ form a subspace of $\mathbb{Z}_2^n$. The rest of the vectors describing the full stabilizer group are constructed using Pauli $Z$ operators. To be precise,
any vector $\vec{v}$ orthogonal to the $X$ subspace can be used to construct the operator $Z^{\vec{v}} = Z^{v_1} \otimes \dots Z^{v_n}$, and it is clear that this string of $Z$'s commutes with every element of $N$ and hence  stabilizes $\ket{\mathcal{N}}$. Let us denote the subgroup generated by the $Z$  strings as $\tilde N$. The direct sum of a subspace of $X$ vectors and its orthogonal complement, the $Z$ vectors, equals the full vector space $\mathbb{Z}_2^n$ thereby showing that $\ket{\mathcal{N}}$ is a stabilizer state with the stabilizer group $G$ which is the direct product $N\times \tilde N$. In fact, for any party $R$, we have $G_{R}=N_{R}\times {\tilde N}_R$. In particular
\begin{align}\label{eq:G-N-tilde-N}
    |G_R|=|N_R||{\tilde N}_R|.
\end{align}

The two-qubit Bell state and the three-qubit GHZ state are X-stabilizer states with $N = \{II, XX\}$ and $N = \{III, XXX\}$, respectively. The remaining stabilizer group generators are obtained by looking for Pauli strings involving only $Z$'s that commute with $N$. Not every stabilizer state is an X-stabilizer state. The generators of a stabilizer state, in general, will involve a mixing of both $X$ and $Z$. In the language of graph states, we can show that any \emph{bipartite} graph is Clifford equivalent to an X-stabilizer state. This can be seen directly from the definition of a graph state in
equation \eqref{eq:GraphStateDef}. For a graph state defined on a bipartite
graph with vertex sets $A$ and $B$, the stabilizer generators take the
form
\begin{equation}
	K_v = X_v \prod_{u \in N(v)} Z_u ,
\end{equation}
where $N(v)$ denotes the neighborhood of vertex $v$. Applying a
Hadamard transformation on all qubits belonging to, say, sublattice
$A$ exchanges $X \leftrightarrow Z$ on those sites, while leaving
operators on sublattice $B$ unchanged. As a consequence, the stabilizer
generators associated with vertices in $A$ are mapped to operators
consisting solely of $Z$’s, whereas those associated with vertices in
$B$ become strings containing only $X$’s. Thus, after the sublattice
Hadamard transformation, the stabilizer group naturally decomposes
into two commuting sets of generators, one purely $Z$-type and the
other purely $X$-type.

The ground states of the toric code \cite{Kitaev:1997wr} and the X-cube model \cite{Haah:2011drr, Vijay:2016phm} are some famous X-stabilizer states. The toric code Hamiltonian is
\begin{equation}
H_{\mathrm{TC}} = -J_e \sum_v A_v \;-\; J_m \sum_p B_p,
\end{equation}
where qubits reside on the links of a two-dimensional square lattice. The \emph{star} and \emph{plaquette} operators are defined as
\begin{equation}
A_v = \bigotimes_{i \in +_v} X_i, \quad B_p = \bigotimes_{i \in \square_p} Z_i,
\end{equation}
where $A_v$ acts on the four edges meeting at the vertex $v$, and $B_p$ acts on the four edges surrounding the plaquette $p$.
All terms in the Hamiltonian commute: $[A_v, B_p] = 0$, and satisfy $(A_v)^2 = (B_p)^2 = 1$.
The ground state obeys $A_v = +1$ and $B_p = +1$ for all $v,p$,
while excitations correspond to deconfined anyons carrying electric ($e$) and magnetic ($m$) charges characteristic of $\mathbb{Z}_2$ topological order.

Similarly, the Hamiltonian of the X-cube model is
\begin{equation}
H_{\mathrm{XC}} = -J_c \sum_{c} A_c \;-\; J_v \sum_{v,\mu} B_v^{\mu},
\end{equation}
where the qubits now reside on the links of a three-dimensional cubic lattice. The operators are defined as
\begin{equation}
A_c = \bigotimes_{i \in \partial c} X_i, \quad B_v^{\mu} = \bigotimes_{i \in \text{plane}_\mu(v)} Z_i,
\end{equation}
where $A_c$ acts on the twelve edges forming the boundary of the cube $c$,
and $B_v^{\mu}$ acts on the four coplanar edges adjacent to a vertex $v$ lying in the plane normal to the direction $\mu \in \{x,y,z\}$.
All terms commute: $[A_c, B_v^{\mu}] = 0$, and satisfy $(A_c)^2 = (B_v^{\mu})^2 = 1$.
The ground state obeys $A_c = +1$ and $B_v^{\mu} = +1$ for all $c,v,\mu$,
while excitations exhibit subdimensional (fractonic) mobility, characteristic of fracton topological order.

The stabilizer group of the toric code is generated by the $A_v$ and $B_p$ operators, while for the X-cube model, it is generated by the $A_c$ and $B_v^{\mu}$ operators. In both cases, the generators split into strings containing only $X$'s and strings containing only $Z$'s. Therefore, the ground states of these models are X-stabilizer states. 

The  R\'enyi entropies of X-stabilizer states were computed in \cite{Hamma:2004vdz}. We present the expression of the $n=2$ Rényi multi-entropy for X-stabilizer states for an arbitrary number of parties $q$. The expression will only involve cardinalities of the group $N$ and its various subgroups.

\begin{figure}
	\centering 
	\includegraphics[width=0.6\linewidth]{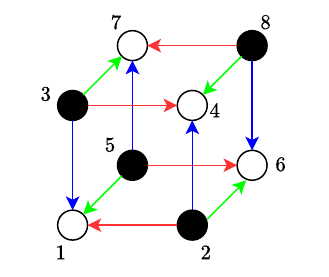}
	\caption{The multi-invariant graph that calculates the $n=2$, $q=3$ R\'enyi multi-entropy, defined in terms of the states. Each black vertex is occupied by $\ket{\psi}$ and each white vertex by $\bra{\psi}$.}
	\label{fig:q=3n=2}
\end{figure}

For concreteness, let us consider a tripartite X-stabilizer state. It is written as
\begin{align}\label{eq:N-state}
    \ket{\mathcal{N}}=\frac{1}{|N|^\frac12}\sum_{g\in N} g|0\rangle^{\otimes n}.
\end{align}
Note the similarity of this form of the state to the expression for the density matrix of a general stabilizer state $|\psi\rangle\langle\psi|=\frac{1}{|G|}\sum_{g\in G} g$. This expression was the basis of our counting argument in section \ref{counting}. Having a similar expression for the state allows us to utilize the same counting but now at the ``state-level'' rather than at the ``density matrix-level''. Consider the ${\cal Z}$ for tripartite $n=2$ R\'enyi multi-entropy. It is given by the graph in figure \ref{fig:q=3n=2}.

We assign a general element $g\in N$ to vertex $1$. As $g^2=1$ for any $g$, we turn on the same $g$ at all vertices. To probe further freedom in assignment, we have to turn on group elements  at other vertices such that $g$ at $1$ is unaffected. This forces us to turn on $g'\in N_{AB}$ at vertices $5,6,7,8$ and so on. We see that this assignment is exactly the one that we used for the doubled graph ${\cal Z}^2$ of density matrices in section \ref{counting}. This gives the expression
\begin{align}\label{eq:N-counting}
    {\cal Z}=\frac{|N_A||N_B||N_C||N_{AB}||N_{BC}||N_{CA}|}{|N|^3}.
\end{align}
Note the similarity with equation \eqref{tripartite-counting}. We have replaced $G$ and its subgroups by $N$ and its corresponding subgroups. In fact, we can make a more direct connection and reproduce equation \eqref{tripartite-counting} from the above formula.

Consider the state \[\frac{1}{|\tilde N|^{1/2}}\sum_{g\in {\tilde N}} g|+\rangle^{\otimes n}.\] Compared to \eqref{eq:N-state}, we have changed $N$ to $\tilde N$ while simultaneously changing the initial state from $|0\rangle$ to $|+\rangle$, the eigenstate of $X$ with eigenvalue $+1$. It is straightforward to see that the stabilizer  group of this state is also $G=N\times {\tilde N}$. As a result, we have an alternate expression for $|N\rangle$,
\begin{align}
    |\mathcal{N}\rangle=\frac{1}{|\tilde N|^{1/2}}\sum_{g\in {\tilde N}} g|+\rangle^{\otimes n}.
\end{align}
The above counting argument goes through unaffected even for this form of the state, giving,
\begin{align}\label{eq:tilde-N-counting}
    {\cal Z}=\frac{|{\tilde N}_A||{\tilde N}_B||{\tilde N}_C||{\tilde N}_{AB}||{\tilde N}_{BC}||{\tilde N}_{CA}|}{|{\tilde N}|^3}.
\end{align}
Multiplying equations \eqref{eq:N-counting} and \eqref{eq:tilde-N-counting} and using \eqref{eq:G-N-tilde-N} gives us the equation \eqref{tripartite-counting}.
Any expression for multi-invariants involving the cardinalities $G$ and its subgroup can be specialized to X-stabilizer states by replacing $G$ with $N$ (or equivalently with $\tilde N$) and taking a square root. In particular, this simplifies the conjecture for Coxeter multi-invariants of X-stabilizer states.

\section{Discussion}\label{sec:discussion}

We conclude with a brief summary and highlight several directions for future research.\\ 

\textbf{Summary.} In this paper, we presented a numerical algorithm to calculate multi-invariants for stabilizer states. We conjectured a formula that computes any Coxeter multi-invariant for a stabilizer state in terms of the stabilizer group and its various subgroups. We verify this conjecture for tripartite stabilizer states. We also show that for X-stabilizer states, the formulas can be further simplified. Although not general, X-stabilizer states describe ground states of interesting many-body Hamiltonians like the toric code and the X-cube model. Our formulas can directly be applied in this context.\\

\textbf{Topology and the multipartite nature of entanglement.} An interesting feature of our work is the role played by topology. In the case of tripartite stabilizer states, we explicitly saw how the exponent of the factor $|G_{AB}\cdot G_{BC} \cdot G_{AC}|$ is the genus $\mathtt{g}$ of the multi-invariant graph. When $\mathtt{g} = 0$, we expect our counting argument to apply as is. However, we don't yet have a handle on the counting argument when the topology is non-trivial.

\begin{figure}
 \centering
 \includegraphics[width = 0.5\linewidth]{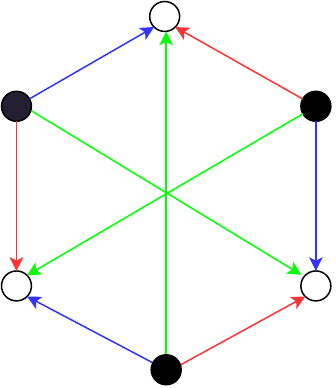}
 \caption{The Kempe invariant. It is the simplest tripartite multi-invariant with a non-trivial topology.}
 \label{fig:kempe-invariant}
\end{figure}

For example, consider the Kempe invariant \cite{Kempe:1999vk} in figure \ref{fig:kempe-invariant}. It is the simplest tripartite multi-invariant with a non-trivial genus, and can easily be computed using the GHZ extraction theorem \ref{ghz-extraction} giving
\begin{equation}
 \mathcal{Z}_{\text{Kempe}}^2 = \frac{|G_A| |G_B| |G_C|}{|G_{AB} \cdot G_{BC} \cdot G_{AC}|}.
\end{equation}
The multi-invariant graph in figure \ref{fig:kempe-invariant} has a non-trivial genus $\mathtt{g} = 1$, resulting in the factor of $|G_{AB}\cdot G_{BC} \cdot G_{AC}|$ above. It would be instructive to obtain this answer using a counting argument. We believe exploring this connection with topology is an interesting avenue for future research. 

Another related feature of the Coxeter multi-invariants is that 
they effectively reduce to bipartite measures of entanglement for stabilizer states. This is because they don't probe the non-trivial subgroup $G_{AB} \cdot G_{BC} \cdot G_{AC}$. In fact, it turns out that any Coxeter multi-invariant $\mathcal{Z}_{\text{cox}}^2$ can be written as a product over bi-partitions as in equation  (\ref{eq:bipartitions}). The non-trivial piece of multiparty entanglement is contained in the genus contribution $|G_{AB} \cdot G_{BC} \cdot G_{AC}|$. In \cite{Fattal:2004frh}, the subgroup $G_{AB} \cdot G_{BC} \cdot G_{AC}$, or more generally, $G_{\hat{A}_1} \cdot \ldots \cdot G_{\hat{A}_q}$, plays a special role as it captures all the generators that localize over $(q-1)$ parties or less. The quantity 
\begin{equation}
	\log_2 \left(\frac{|G|}{|G_{\hat{A}_1} \cdot G_{\hat{A}_2} \cdot \ldots \cdot G_{\hat{A}_q}|} \right),
\end{equation}
therefore, captures genuine multiparty entanglement in stabilizer states \cite{Fattal:2004frh}. The Coxeter multi-invariants for stabilizer states, in this sense, do not probe genuine multipartite entanglement. 
\\

\textbf{Entanglement phase transitions.} Random quantum circuits are being studied extensively \cite{Fisher_etal_review_2023}. A special class of them are the Clifford circuits. Clifford unitary circuits with local Pauli measurements demonstrate a remarkable entanglement phase transition from a volume-law to an area-law as the rate of measurements is increased \cite{Li:2018mcv, Skinner:2018tjl,Liu_etal_2022,Guo_etal_2023}. The unitaries create entanglement while the Pauli measurements  destroy entanglement. By tuning the frequency of measurements, there is a continuous phase transition from volume-law to area-law entanglement.

Our formulas for multi-invariants are directly applicable to time evolution generated by Clifford unitaries with interspersed Pauli measurements since all these operations are within  the stabilizer framework. An interesting future direction is to study multiparty entanglement phase transitions in these systems. For example, \cite{Zabalo:2019sfl} looks at the tripartite mutual information in such models. However, the tripartite mutual information is simply a combination of bipartite entanglement entropies, and does not truly capture properties of genuine multiparty entanglement. A few more works that look at multiparty entanglement phase transitions in these models are \cite{Avakian:2024xiv, Paviglianiti:2023gjy, DiFresco:2023mjh}. We leave the study of entanglement phase transitions using multi-invariants in these systems to future work. \\

\textbf{Fractons.} Fracton quantum stabilizer codes like Haah's code or the X-cube model have a very rich entanglement structure and excitation mobility. Our results are applicable to these kind of systems and might offer new physical insights. We leave explorations in this direction to future work.

\begin{acknowledgments}
We thank Gaurav Aggarwal, Bartek Czech, Kedar Damle, Shraiyance Jain, Onkar Parrikar, and Shiroman Prakash for useful discussions. Part of this work was presented at the International Centre for Theoretical Sciences (ICTS) during the program - Quantum Information, Quantum Field Theory and Gravity (code: ICTS/qftg2024/08). This work is supported by the Department of Atomic Energy, Government of India, under Project Identification No. RTI 4002, and the
Infosys Endowment for the study of the Quantum Structure of Spacetime.
\end{acknowledgments}

\bibliography{references}

\appendix

\section{Counting for tripartite graph states}\label{graph_state_counting}

We work with the graph state related to $\ket{\psi}$ by a LU transformation. If $\gamma$ is the adjacency matrix of this graph, then the stabilizer group $G$ is generated by the elements:
\begin{equation}
	g_a = X_a \prod_{b} Z_b^{\gamma_{ab}}
\end{equation}
Each element of the stabilizer group can therefore be labelled by a binary vector $\vec{u} \in \mathbb{Z}_2^n$, where $n$ is the total number of qubits in the state $\ket{\psi}$. We will denote the corresponding stabilizer group element by $g(\vec{u})$ which is defined as
\begin{equation}
	g({\vec{u}}) = \prod_{a: u_a = 1} g_a.
\end{equation}
If we forget about some minus signs coming from the commutation of $X$ and $Z$, this stabilizer group element can be thought of as
\begin{equation}
	g(\vec{u}) \sim X^{\vec{u}} Z^{\gamma \cdot \vec{u}}
\end{equation}
with the notation $X^{\vec{u}} = X_1^{u_1} \otimes X_2^{u_2} \dots X_n^{u_n}$, and similarly for $Z^{\vec{u}}$. It'll be useful to think of these vectors as made up of three constituent vectors: $\vec{u} = (u_A, u_B, u_C)$. Ignoring the minus signs, we then have
\begin{equation}
	g(\vec{u}) \sim (X^{u_A} Z^{(\gamma \cdot \vec{u})_A})\otimes (X^{u_B} Z^{(\gamma \cdot \vec{u})_B}) \otimes (X^{u_C} Z^{(\gamma \cdot \vec{u})_C}).
\end{equation}

The multi-invariant in figure \ref{fig:cube-grid} becomes a product of traces. In its full glory, it is
\begin{widetext}
\begin{equation}
	\begin{split}
	\mathcal{Z}_{n=2}^{\text{multi-entropy}} = \frac{1}{|G|^8} \sum_{\sigma_1, \dots, \sigma_8 \in G} & \Tr(\sigma_{1A} \sigma_{2A}) \Tr(\sigma_{3A} \sigma_{4A}) \Tr(\sigma_{5A} \sigma_{6A}) \Tr(\sigma_{7A} \sigma_{8A}) \\
	\times & \Tr(\sigma_{1B} \sigma_{3B}) \Tr(\sigma_{2B} \sigma_{4B}) \Tr(\sigma_{5B} \sigma_{7B}) \Tr(\sigma_{6B} \sigma_{8B}) \\
	\times & \Tr(\sigma_{1C} \sigma_{5C}) \Tr(\sigma_{2C} \sigma_{6C}) \Tr(\sigma_{3C} \sigma_{7C}) \Tr(\sigma_{4C} \sigma_{8C}),
	\end{split}
\end{equation}
\end{widetext}
with the notation $\sigma_{1} = \sigma_{1A} \otimes \sigma_{1B} \otimes \sigma_{1C}$, etc. There's a dramatic simplification that occurs because the stabilizer group is made of Pauli operators that are traceless. The only contribution to the trace comes from the identity. Furthermore, each stabilizer group element squares to the identity, therefore this gives us various constraints on the $\sigma$'s that contribute. For example, from the first trace, we get $\sigma_{2A} = \sigma_{1A}$.

\begin{figure}
	\includegraphics[width = 0.75\linewidth]{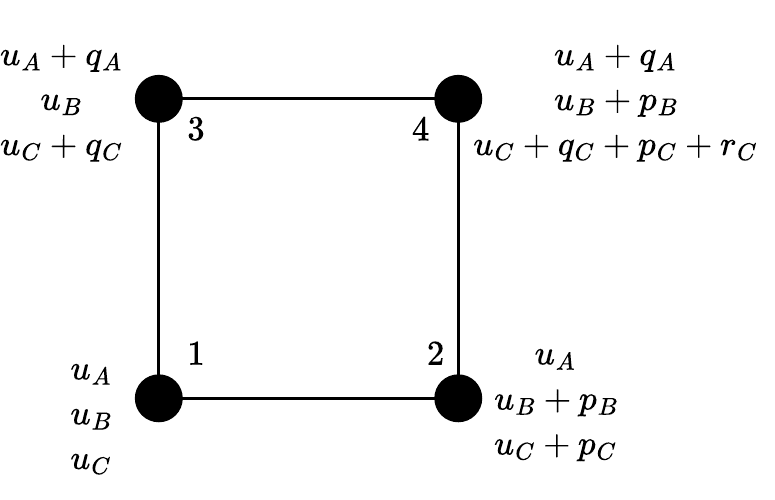}
	\caption{Assignment of vectors to the four vertices labeled by $1$, $2$, $3$, $4$, of the $q=3$, $n=2$ multi-entropy grid from figure \ref{fig:q=3n=2}.}
	\label{fig:1234-square}
\end{figure}

To simplify the sum above, we need to count the solutions of the set of constraints coming from the expression for the multi-entropy above. To do this, let us first fix the vector at site $1$ to be $\vec{u} = (u_A, u_B, u_C)$. There are $|G|$ ways to pick this vector $\vec{u}$. We will choose the other vectors as shown in figure \ref{fig:1234-square} for sites $2$, $3$, and $4$.

\begin{figure}
	\includegraphics[width = \linewidth]{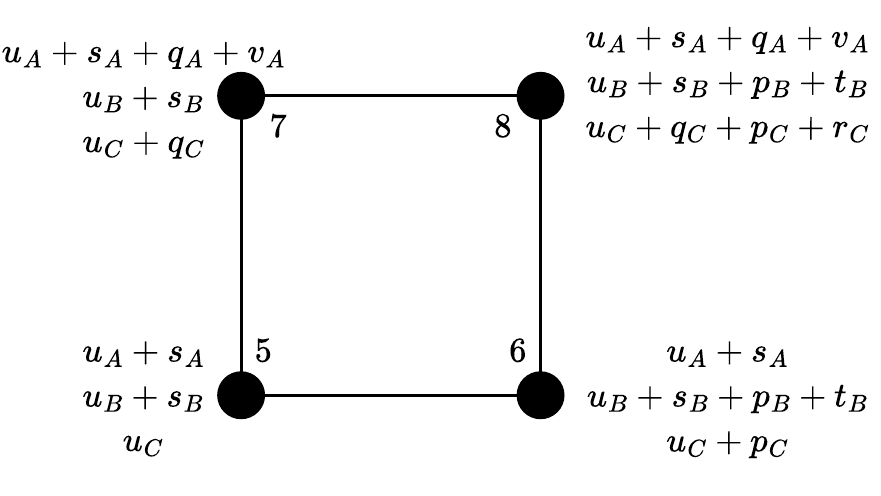}
	\caption{Assignment of vectors to the vertices labeled by $5$, $6$, $7$, $8$, of the $q=3$, $n=2$ multi-entropy grid from figure \ref{fig:q=3n=2}.}
	\label{fig:5678-square}
\end{figure}

The constraints on these vectors are:
\begin{equation}
	\begin{split}
		& \gamma_{AB} p_B + \gamma_{AC}p_C  = 0 \\
		& \gamma_{BA}q_A + \gamma_{BC}q_C  = 0 \\
		& \gamma_{BC}r_C  = 0\\
		& \gamma_{AC} r_C  = 0.
	\end{split}
\end{equation}
From this we conclude that $p \in G_{BC}$, $q \in G_{AC}$, and $r \in G_C$. Similarly, we fix the vectors at the sites $5$, $6$, $7$, and $8$ as shown in figure \ref{fig:5678-square}. The constraints on these vectors take the form:
\begin{equation}
	\begin{split}
		& \gamma_{CA}s_A + \gamma_{CB}s_B = 0 \\
		& \gamma_{CB}t_B = 0 \\
		& \gamma_{AB}t_B = 0 \\
		& \gamma_{BA} v_A = 0 \\
		& \gamma_{CA} v_A = 0.
	\end{split}
\end{equation}
From these equations, we conclude that $s \in G_{AB}$, $t \in G_B$, and $v \in G_A$.

The counting goes as follows. There are $|G|$ ways to pick $\vec{u}$, $|G_{BC}|$ ways to pick $\vec{p}$, $|G_{AC}|$ ways to pick $\vec{q}$, $|G_C|$ ways to pick $\vec{r}$, $|G_{AB}|$ ways to pick $\vec{s}$, $|G_B|$ ways to pick $t$, and finally $|G_A|$ ways to pick $\vec{v}$. Putting all of this together, we get back equation  (\ref{tripartite-counting}), 
\begin{equation}
	\mathcal{Z} = \frac{1}{|G|^3} |G_A||G_B||G_C||G_{AB}||G_{BC}||G_{AC}|.
\end{equation}

\section{Another efficient algorithm}\label{another}
In this appendix, we review the inner product algorithm of \cite{garcia2013efficientinnerproductalgorithmstabilizer}.
Given an $n$-qubit stabilizer state $\ket{\psi}$, we have the generators $\{g_1, \dots, g_n\}$ which generate the stabilizer group of $\ket{\psi}$. Each generator is a Pauli string of $n$ Pauli matrices plus a phase picked from $\{\pm 1, \pm i\}$. We can therefore arrange these $n$ strings in the form of an $n \times n$ matrix.

As an example, consider the state $\ket{\psi} = \left( \ket{00} + \ket{11}\right)/\sqrt{2}$. It is stabilized by $XX$ and $ZZ$, and its stabilizer matrix is
\begin{equation}
	\mathcal{M}_1 = \begin{vmatrix}
		X X \\ Z Z
	\end{vmatrix}.
\end{equation}
We could choose a different set of generators, say, $\{XX, -YY\}$ or $\{-YY, ZZ\}$ that generate the same stabilizer group of the state $\ket{\psi}$, and they are represented as
\begin{equation}
	\mathcal{M}_2 = \begin{vmatrix}
		XX \\ - YY
	\end{vmatrix}, \quad \text{and} \quad \mathcal{M}_3 = \begin{vmatrix}
		-YY \\ ZZ
	\end{vmatrix}.
\end{equation}

\begin{figure}
 \centering
 \includegraphics[width = 0.75\linewidth]{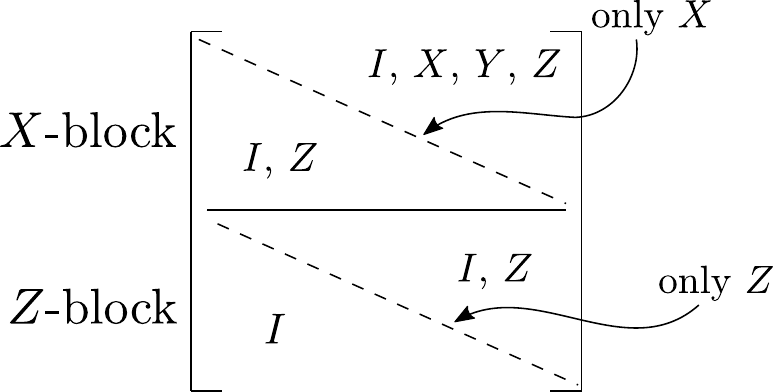}
 \caption{Canonical form of the stabilizer matrix.}
 \label{fig:canonical-form}
\end{figure}

Given a stabilizer matrix, it is easy to see that other stabilizer matrices of the same state are generated by: (a) interchanging rows, (b) multiplying two rows. For example, $\mathcal{M}_2$ is obtained from $\mathcal{M}_1$ by multiplying the second row with the first. Similarly $\mathcal{M}_3$ is obtained from $\mathcal{M}_1$ by multiplying the first row by the second. Using these two operations, it is possible to bring the stabilizer matrix to the canonical form \cite{Aaronson:2004xuh} shown in figure \ref{fig:canonical-form}.

Given an $n$-qubit stabilizer state $\ket{\psi}$, it is possible to use only CNOT, Hadamard, and phase gates to bring it to the state $\ket{0}^{\otimes n}$. In other words,
\begin{equation}
 \ket{\psi} = U \ket{0}^{\otimes n}
\end{equation}
where $U$ consists entirely of CNOT, Hadamard, and phase gates \cite{Aaronson:2004xuh}, and $\ket{0}$ is the eigenvector of $Z$ with eigenvalue $+1$. The stabilizer matrix for the state $\ket{0}^{\otimes n}$ is
\begin{equation}
 \mathcal{M} = \begin{vmatrix}
  Z & I & \dots & I \\
  I & Z & \dots & I \\
  \vdots & \vdots & \ddots & \vdots \\
  I & I & \dots & Z
 \end{vmatrix}.
\end{equation}
The following results are key to the inner product algorithm of \cite{garcia2013efficientinnerproductalgorithmstabilizer}.

First, if $\ket{\psi}$ is a stabilizer state, then any element $P$ belonging to its stabilizer group has the form:
\begin{equation}
 P = i^k P_1 \otimes \dots \otimes P_n
\end{equation}
where the phase $k$ is either $0$ or $2$ and each $P_i$ is either $I$, $X$, $Y$, or $Z$. The reason is because $P^2$ should also stabilize $\ket{\psi}$, and this cannot happen if $k = 1$ or $3$.

Second, if $\ket{\psi}$ and $\ket{\phi}$ are two stabilizer states and there is a Pauli operator $P$ in the stabilizer group of $\ket{\psi}$ such that $P \ket{\phi} = - \ket{\phi}$, then $\langle \phi | \psi\rangle = 0$. The proof is easy:
\begin{equation}
 \langle\phi|\psi\rangle = \langle \phi| P |\psi \rangle = - \langle \phi|\psi \rangle,
\end{equation}
where we used $P\ket{\psi} = \ket{\psi}$ in the first equality and $P\ket{\phi} = -\ket{\phi}$ in the second equality.

Third, if $S(\psi)$ and $S(\phi)$ are the stabilizer groups of two non-orthogonal states $\ket{\psi}$ and $\ket{\phi}$, then
\begin{equation}\label{eq:inp}
 |\langle\phi|\psi\rangle| = 2^{-s/2}
\end{equation}
where $s = n - r$, and $r$ is the number of independent generators of the group $S(\psi) \cap S(\phi)$. To see this, consider the square of the inner product
\begin{equation}
 |\langle \phi | \psi \rangle|^2 = \frac{1}{2^{2n}} \sum_{\sigma \in S(\psi)} \sum_{\tau \in S(\phi)} \Tr (\sigma \tau),
\end{equation}
where we used equation  (\ref{eq:density}). For a fixed $\sigma$, the trace is non-zero only when $\tau = \pm \sigma$. By assumption, $\tau \neq - \sigma$ since the inner product will be zero otherwise. Hence $\tau = \sigma$ and the sum only clicks for elements in the intersection $S(\psi) \cap S(\phi)$. If this group has $r$ generators, then there are $2^r$ elements that contribute to the sum above. Since the trace of the identity is $2^n$, we get equation  (\ref{eq:inp}).

The inner product algorithm goes as follows. First, find a unitary circuit $U$ that takes a state $\ket{\psi}$ to the basis form $\ket{0}^{\otimes n}$. Second, find the canonical form of the stabilizer matrix of the state $U^{\dagger} \ket{\phi}$. Then $s$ is the number of rows belonging to the $X$-block of this stabilizer matrix.

The complexity of this algorithm and complexity of our algorithm is compared in figure \ref{fig:time-complexity-python}.
\end{document}